\documentclass[11pt]{article}

\usepackage[preprint]{acl}

\usepackage{times}
\usepackage{latexsym}

\usepackage[T1]{fontenc}

\usepackage[utf8]{inputenc}

\usepackage{microtype}

\usepackage{inconsolata}

\usepackage{graphicx}

\usepackage{booktabs}
\usepackage{multirow}
\usepackage{array}
\usepackage{makecell}
\usepackage{subcaption}
\usepackage{tabularx}
\usepackage{longtable}
\usepackage{adjustbox}
\usepackage{siunitx}
\usepackage{pifont}
\usepackage{amsmath}
\usepackage{amssymb}
\usepackage{titling}
\title{Mechanism-Level Evaluation for Vision-Language Models: Controlled Activation-Replacement Diagnosis of Gender Bias}
\author{
  \textbf{Zhipeng Zhao}\textsuperscript{1},
  \textbf{Wenxu Wang}\textsuperscript{1},
  \textbf{Peishun Liu}\textsuperscript{1},
  \textbf{Ruichun Tang}\textsuperscript{1,*}
\\
\\
  \textsuperscript{1}Ocean University of China
\\
  \small{\texttt{\{zhaozhipeng\}@stu.ouc.edu.cn}} \\
  \small{\texttt{\{wangwenxu, liups, tangruichun\}@ouc.edu.cn}}
\\
  \small{\textsuperscript{*}Corresponding author}
}

\begin{document}
\maketitle
\begin{abstract}
Behavioral benchmarking reveals \emph{what} biases exist in vision-language models but not \emph{which internal components} are most sensitive to targeted intervention, precluding principled intervention. We argue for mechanism-level evaluation as a necessary complement, demonstrating causal mediation analysis as a diagnostic instrument for gender bias. We decompose gender-cue effects into controlled indirect effects attributable to specific-layer activations and direct effects through all other pathways, producing layer-by-layer mechanistic signatures. Across six models spanning three architectural families (LLaVA-1.5, LLaVA-NeXT, InstructBLIP at 7B/13B) and two 8B-scale architectures, three findings emerge: language-layer activations exhibit the greatest output sensitivity under controlled intervention, with the direct component often carrying the opposite sign; architectural choices redistribute layer-wise sensitivity to activation replacement; and counterfactual scores diverge from surface-level scores, exposing implicit associations. Systematic ablation validates internal consistency. An intervention experiment finds that the average indirect effect (AIE) and downstream intervention effectiveness are only weakly correlated (Pearson $r = 0.33$), and the layer with the second-largest AIE produces near-zero bias change---indicating that mechanistic diagnosis captures activation-replacement sensitivity but does not, by itself, identify optimal intervention targets. These results show mechanism-level evaluation captures architecture-specific sensitivity patterns that behavioral benchmarks cannot; pairing both should become standard NLP practice. Code: \url{https://github.com/zhaozhipeng1997/CARD-GenderBias}.
\end{abstract}

\section{Introduction}

Vision-language models (VLMs) have shifted from research prototypes to deployed infrastructure, yet evaluation methods remain largely unchanged from an era of narrow models. This prompts a question: \textit{What should NLP evaluation look like?}

This paper offers one answer: evaluation must move beyond behavioral benchmarking to mechanism-level diagnosis. Behavioral benchmarks measure \emph{what} a model outputs; extensive work documents systematic biases in VLMs~\citep{xiao2024genderbiasemphvlbenchmarkinggenderbias,howard-etal-2025-uncovering,girrbach2025revealingreducinggenderbiases}. Yet these treat the model as a black box---recording outputs without revealing \emph{which internal components} are most sensitive to targeted intervention. Where does activation replacement produce the largest output shift---vision encoder, language decoder, or cross-modal connector? Debiasing methods targeting the wrong component may be ineffective, yet behavioral evaluation offers no guidance.

Mechanism-level evaluation addresses this gap: intervene on a specific internal component, measure the output change, and attribute the change to that component---a logic underlying causal inference and mechanistic interpretability that should complement behavioral benchmarking.

We demonstrate this through causal mediation analysis~\citep{mackinnon2007mediation}, introduced to NLP by \citet{vig2020causalmediationanalysisinterpreting}. It decomposes a total treatment effect into a controlled indirect effect (AIE) through a specific layer's activations and a direct effect (ADE) through all other pathways. For each layer, we manipulate the gender cue (treatment), collect that layer's activations (mediator), measure confidence in a gendered occupation response (outcome), and control the mediator at baseline-gender values under reversed-gender prompts. Repeating across every layer yields a \emph{mechanistic signature}---a layer-by-layer map of where activation replacement most strongly shifts the output.

We apply this to six models spanning three families---LLaVA-1.5, LLaVA-NeXT, and InstructBLIP (7B/13B)---plus MiniCPM-V~2.6 and \mbox{InternVL3.5} (8B), on FACET and MS COCO. Three findings emerge. First, \textbf{language-layer activations exhibit the greatest output sensitivity}, with the direct component often carrying the opposite sign. Second, \textbf{architectural choices redistribute which layers are most sensitive}: LLaVA-NeXT shows elevated vision-layer sensitivity; InstructBLIP's Q-Former concentrates sensitivity in language layers under occupation prompts---a pattern that inverts when prompts omit occupation (§\ref{subsec:multi_attribute}). Third, \textbf{counterfactual scores systematically diverge from factual scores}, exposing implicit associations. Systematic ablation validates diagnostic consistency; an intervention experiment reveals a weak correlation between diagnostic sensitivity and downstream intervention effectiveness, cautioning that the diagnostic signal does not directly translate to intervention targets.

We contribute: (1) arguing that mechanism-level evaluation should complement behavioral benchmarking; (2) applying causal mediation to six VLMs, producing architecture-specific mechanistic signatures; (3) through systematic ablation and intervention validation, providing a reusable methodological template and demonstrating that AIE does not straightforwardly translate to intervention effectiveness---cautioning against using signatures as targets without separate validation. Code and data are public.
\section{Related Work}
\label{sec:related_work}
\begin{figure*}[ht]
	\centering
	\includegraphics[width=\textwidth]{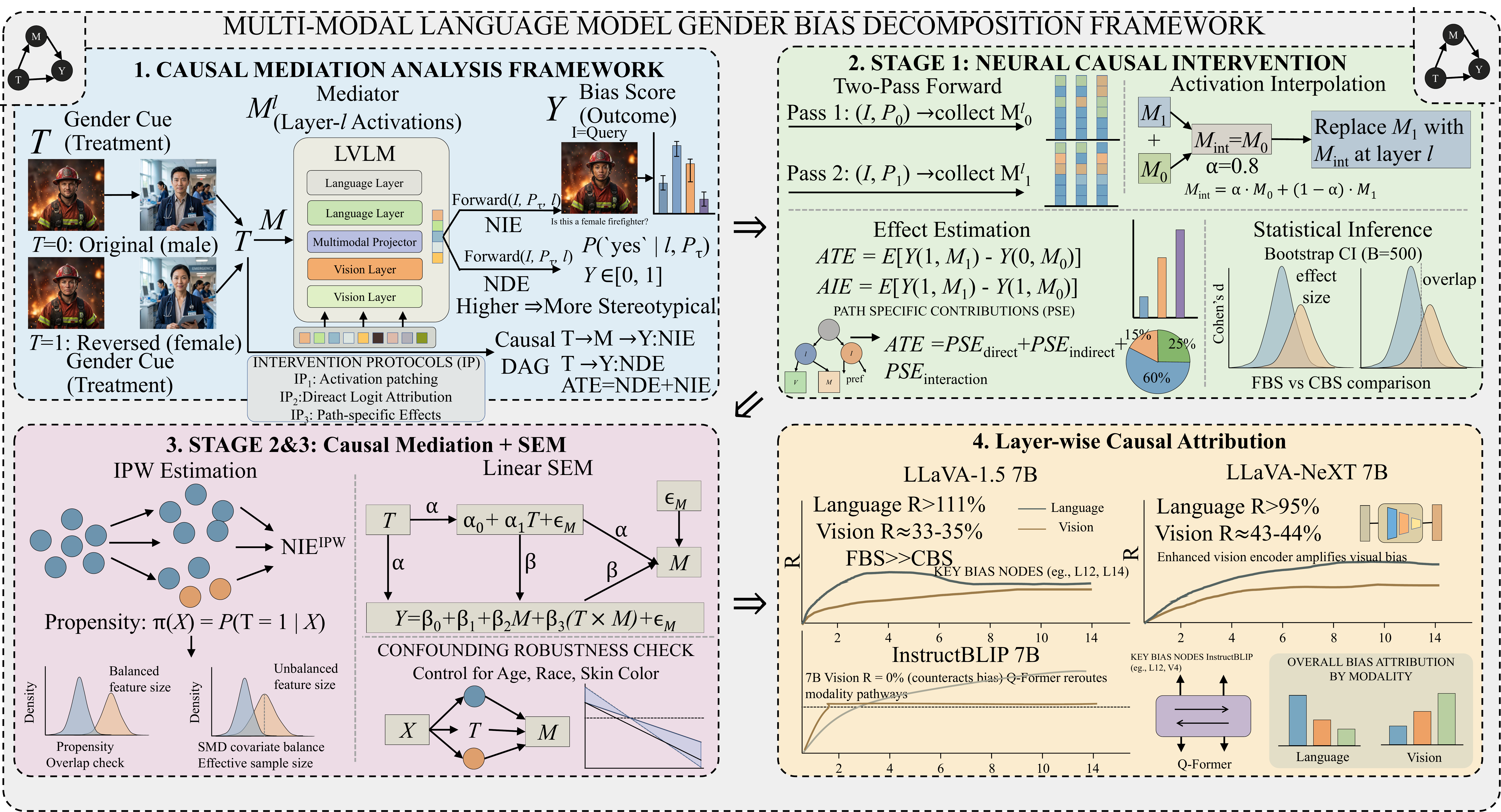}
	\caption{\textbf{Activation-replacement sensitivity framework:} (1) experimental manipulation via neural intervention, (2) diagnostic checks via IPW and SEM, (3) sensitivity analysis.}
	\label{fig:method_framework}
\end{figure*}
\subsection{The Boundary of Behavioral Benchmarking}
\label{subsec:rw_measurement}

The dominant paradigm treats models as black boxes---inputs in, outputs recorded, biases quantified~\citep{xiao2024genderbiasemphvlbenchmarkinggenderbias,howard-etal-2025-uncovering,janghorbani-de-melo-2023-multi}. Benchmarks document gender bias in VLMs with growing rigor, yet share a structural boundary: they establish \emph{that} bias exists, not \emph{which components} are most sensitive to targeted intervention.

\citet{zhou-etal-2022-vlstereoset} introduced VLStereoSet (1,028 image--caption pairs); \citet{ruggeri-nozza-2023-multi} found 5\% toxicity in ViLT/VisualBERT completions; \citet{cabello-etal-2023-evaluating} showed external fairness metrics cannot diagnose internal bias loci; \citet{hall-etal-2023-vision} and \citet{fraser-kiritchenko-2024-examining} analyzed gender disparities across VLM zero-shot and instruction-tuned settings. At scale, \citet{xiao2024genderbiasemphvlbenchmarkinggenderbias} (34,581 VQA pairs), \citet{howard-etal-2025-uncovering} (171K images), and \citet{girrbach2025revealingreducinggenderbiases} (22 assistants) convergently show gender bias is pervasive and architecture-dependent. \citet{sathe-etal-2024-unified} and \citet{abdollahi-etal-2024-gabinsight} propose unified benchmarks for gender-profession and gender-activity bias; \citet{janghorbani-de-melo-2023-multi} introduced MMBias, covering 14 population subgroups beyond gender and race. \citet{vo2025vlmsbiased} showed VLMs are strongly biased by Internet priors, ignoring visual counter-evidence---reinforcing the need for internal diagnostics. Our multi-attribute analysis (§\ref{subsec:multi_attribute}) examines whether mechanistic bias pathways are modulated by skin tone and hair type.

These efforts represent the state of the art in behavioral documentation, yet cannot answer \emph{which components} are most sensitive under controlled intervention---they do not observe internal representations. Both evaluation modes are necessary; neither is sufficient.

\subsection{Causal Analysis as Scientific Instrumentation}
\label{subsec:rw_causal}

A growing line of work treats neural models as scientific objects. Causal mediation analysis~\citep{mackinnon2007mediation} decomposes treatment effects into contributions from specific internal variables. \citet{vig2020causalmediationanalysisinterpreting} introduced this in NLP, finding gender bias in GPT-2 is sparse ($\sim$4\% of neurons). We extend from text-only to VLMs and from component-level to layer-level.

\citet{weng2025imagesspeaklouderwords} applied causal mediation to VLMs at module granularity; we decompose at the layer level within each modality. \citet{kavuri-etal-2025-freeze} proposed modality-freezing; we additionally quantify per-layer contributions. \citet{palit2023towardsvisionlanguagemechanistic} used causal tracing in BLIP; \citet{neo2025interpretingvisiongrounding} analyzed layer-wise visual grounding; \citet{shahgir2026vlmsneedwords} showed VLMs ignore visual detail for semantic anchors. At finer granularity, \citet{yasser2026locating} localize demographic bias at individual attention heads in CLIP's vision encoder; we complement this by diagnosing bias across every layer in generative VLMs with an interventional estimator, covering both modalities. For text-only models, \citet{todd-etal-2023-function} demonstrated localized functional interventions; \citet{chintam-etal-2023-identifying} connected gender bias diagnosis to targeted component adaptation.

On mitigation, \citet{zhang-etal-2024-think} (GAMA), \citet{an2026interpretabledebiasingvisionlanguagemodels} (DEBIASLENS), and \citet{lan2026unveilingfairnessseesawdiscovering} (RES-FAIR) achieve debiasing without layer-level diagnosis; our framework supplies this diagnostic capability. \citet{lin2025biopro} propose training-free subspace projection for difference-aware gender debiasing---while our signatures could inform such projection targets, our intervention experiment ({\S}\ref{subsec:intervention_validation}) shows the diagnostic-to-intervention link is not straightforward. \citet{fu2026diagnosingrepairingunsafe} use causal discovery to localize harmful VLM pathways. Closest in spirit, \citet{wang-etal-2025-vseam} use activation editing; \citet{liu2026seeingnotbelieving} probe visual attention under intervention. We differ in decomposing across every layer with systematic validation. Complementarily, \citet{golovanevsky2024whatvlmsnotice} study mechanistic VLM pipelines, \citet{schaumloffel2026mechanismsobject} study layer-wise object localization, \citet{lin-etal-2025-survey} survey mechanistic interpretability for multi-modal models, \citet{thube2025pathological} analyze VLM failure modes causally, and \citet{conmy-etal-2023-towards} scale causal-intervention analysis. Methodologically, \citet{han2024beyondsurface} assess LLM comprehension via causal effects beyond behavioral scores; \citet{farahani2024decipheringinterplay} apply mediation to internal representations in modern LMs.

Taken together, behavioral benchmarks document \emph{what} biases exist; this paper supplies layer-wise activation-replacement sensitivity analysis applied and validated across architectures. We align with \citet{bouchouchi2026unifiedframework} who argue internal and behavioral evaluations should be paired, and \citet{anonymous2025moralsensitivity} who combine behavioral profiling with mechanistic analysis. On mediator design, \citet{mueller-etal-2024-quest} survey mediator selection; \citet{geiger-etal-2023-causal} examine when internal representations are appropriate explanatory units.

\section{Activation-Replacement Sensitivity Analysis}\label{sec:method}

Causal mediation provides the decomposition framework, but our implementation measures output change under controlled activation replacement rather than identifying natural causal pathways. Probing, patching, and circuit discovery serve related roles; we adopt mediation decomposition because it cleanly separates effects into layer-wise signatures.

\subsection{Problem Definition}\label{subsec:problem_definition}

Given model $\mathcal{M}$, image $I$, occupation $O$: treatment $T$ is gender-cue manipulation ($T{=}1$ for reversed gender, $T{=}0$ for original); mediator $M$ is the scalar activation at a specific layer; outcome $Y$ is the ``yes'' probability to a gendered occupation prompt. We decompose the average total effect (ATE) into the average indirect effect (AIE) attributable to that layer under controlled intervention, and the direct effect (ADE = ATE $-$ AIE) through all other pathways. Repeating across layers yields a \emph{mechanistic signature}.

\subsection{Intervention Design and Effect Estimation}
\label{subsec:our_approach}

The framework uses neural intervention to estimate ATE and AIE. Inverse probability weighting (IPW) and structural equation modeling (SEM) are included as diagnostic transparency checks, not as independent causal identification.

\subsubsection{Intervention Design}
\label{subsubsec:intervention_design}

\textbf{Treatment}: $T{=}0$ uses the true gender $G\_{\text{true}}$ (``Is there a \{$G\_{\text{true}}$\} \{occupation\} in the photo?''); $T{=}1$ uses the opposite gender $G\_{\text{reverse}}$. \textbf{Mediator}: $M$ is the scalar mediator derived from activations at layer $l$ (see \S\ref{subsubsec:mediator_construction}), covering visual layers ($v{:}i$), language layers ($l{:}i$), and the multimodal projector. \textbf{Outcome}: $Y = P(\text{``yes''} \mid I, P_{\text{prompt}})$.

\subsubsection{Mediator Construction}\label{subsubsec:mediator_construction}
High-dimensional layer activations are reduced to a scalar. We extract the last-token hidden state $\mathbf{h}_l \in \mathbb{R}^{d}$ from layer $l$ during forward propagation. PCA is fit separately for each layer, model, dataset, and prompt pair on all treatment-conditioned samples, using flattened $\mathbf{h}_l$ vectors. The PC1 direction $\mathbf{v}_{\text{pc1}}$ is retained; the scalar mediator is $M = \langle \mathbf{h}_l, \mathbf{v}_{\text{pc1}} \rangle$.

PC1 captures maximal treatment-conditional variance. Table~\ref{tab:mediator_summary} confirms L2 norm and mean pooling discard treatment-relevant information, yielding near-zero AIE. We acknowledge PC1 selects the projection most accentuating treatment differences, which could amplify apparent mediation. Table~\ref{tab:mediator_robustness} confirms the qualitative pattern (shallow near zero, deep dominant) is stable across same-data PC1, held-out PC1, and difference-of-means, though absolute ratios vary ($l$:31: 58.4\%--96.3\%). See Limitations.

\subsubsection{Neural Intervention}
\label{subsubsec:neural_intervention}

\textbf{What is intervened on.} The intervention replaces the full activation tensor at the target layer, not the scalar mediator (which only defines the direction of maximal treatment-conditional variance). Given a forward pass with reversed-gender prompt $P_1$, we collect the full activation $f_l(I, P_1) \in \mathbb{R}^{n \times d}$ at layer $l$, replace it with an interpolated activation, and continue the forward pass---analogous to activation patching with interpolation controlled by the scalar mediator direction.

\textbf{Procedure.} For input $(I, P)$ at target layer $l$: collect $\mathbf{A}_0 = f_l(I, P_0)$ (original-gender activations) and $\mathbf{A}_1 = f_l(I, P_1)$ (reversed-gender activations). The counterfactual forward pass uses $\mathbf{A}_{\text{intervened}} = \alpha \cdot \mathbf{A}_0 + (1{-}\alpha) \cdot \mathbf{A}_1$, $\alpha \in [0,1]$ ($\alpha{=}1$: full replacement). Cosine similarity $\geq$0.9984 and KL divergence ${<}0.15$ confirm the intervened activation stays on the manifold. Default $\alpha{=}0.4$; ablation (§\ref{subsec:ablation}, Fig.~\ref{fig:strength_ablation}) confirms qualitative robustness.

\subsubsection{Causal Effect Estimation}
\label{subsubsec:causal_effect_estimation}

$\text{ATE} = \mathbb{E}[Y(1, M(1)) - Y(0, M(0))]$, $\text{AIE} = \mathbb{E}[Y(1, M(1)) - Y(1, M(0))]$, $\text{ADE} = \text{ATE} - \text{AIE}$. Contribution ratio $R_{\text{contrib}} = \text{AIE}/\text{ATE} \times 100\%$; opposite signs can push this beyond 100\% or below 0\%---arithmetic, not compensatory. 95\% bootstrap CIs ($B{=}500$) and Cohen's $d$ accompany all estimates.

These are \emph{controlled} effects: output change under forced replacement, not spontaneous (natural) change. The natural indirect effect is near zero (§\ref{subsec:ablation}, Table~\ref{tab:estimator_comparison})---the model \emph{can} represent gender information without deploying it. \textbf{Sign convention:} negative AIE = intervention reduces ``yes'' probability; positive = increase.

\subsubsection{Diagnostic Robustness Checks}
\label{subsubsec:causal_mediation_analysis}

We use IPW and SEM as diagnostic transparency checks for the observational (natural) mediation estimands only. \textbf{Notation:} NIE/NDE denote natural indirect/direct effects from observational estimators; AIE/ADE denote controlled indirect/direct effects from our interventional method. These are distinct quantities and should not be conflated.

\textbf{IPW}: We apply IPW to verify mediator overlap. The propensity score $\pi(X) = P(T{=}1 \mid X)$; since treatment is assigned by prompt construction, covariates are constant, so IPW verifies overlap sufficiency rather than adjusting for confounding:
\begin{equation}
\begin{aligned}
\widehat{\text{NIE}}^{\text{IPW}} &= \frac{1}{n}\sum_{i=1}^n \left[ \frac{T_i Y_i}{\hat{\pi}} - \frac{T_i Y_i(1,M_i(0))}{\hat{\pi}} \right],\\
\widehat{\text{NDE}}^{\text{IPW}} &= \frac{1}{n}\sum_{i=1}^n \left[ \frac{T_i Y_i(1,M_i(0))}{\hat{\pi}} - \frac{(1-T_i)Y_i}{\hat{\pi}} \right].
\end{aligned}
\end{equation}

\textbf{SEM}: We fit a linear structural equation model: $M = \alpha_0 + \alpha_1 T + \epsilon_M$, $Y = \beta_0 + \beta_1 T + \beta_2 M + \beta_3 (T \times M) + \epsilon_Y$. This tests whether high-dimensional activation-to-outcome relationships can be captured by linear parametric decomposition; the results are predominantly negative (Section~\ref{sec:experiments}). Under linearity: $\text{NDE} = \beta_1 + \beta_3 \mathbb{E}[M(0)]$, $\text{NIE} = \alpha_1 \beta_2$, $\text{ATE} = \text{NDE} + \text{NIE}$.

\subsubsection{Sensitivity Analysis}
\label{subsubsec:sensitivity_analysis}

For confounding strength $\gamma$ and correlation $\rho$, adjusted effects are $\text{ATE}_{\text{adj}} = \widehat{\text{ATE}} - \gamma\rho$, $\text{NIE}_{\text{adj}} = \widehat{\text{NIE}} - 0.5\gamma\rho$. The robustness score $R_{\text{robust}}$ quantifies the proportion of confounding scenarios under which conclusions hold.

\subsubsection{Evaluation Metrics}
\label{subsubsec:evaluation_metrics}

$\text{FBS} = \frac{1}{n_1}\sum_{i: T_i=1} Y_i$ is the global model-level behavioral bias score. $\text{CBS} = \frac{1}{n}\sum_{i=1}^n Y_i(1, M_i(0))$ is computed from counterfactual output probabilities under controlled mediator replacement. FBS and CBS operate in probability space ($[0,1]$); AIE and ATE are in logit-difference space (§\ref{subsec:ablation}, Appendix Table~\ref{tab:outcome_definition}), so their numerical differences are not directly comparable. In result tables, FBS and CBS are constant across rows per model-dataset pair.

\section{Experiments}
\label{sec:experiments}

We test: Can the instrument distinguish architectures by where activation replacement shifts output? (§\ref{subsec:main_results}) Is it reliable? (§\ref{subsec:ablation}) Robust to nuisance variation? (§\ref{subsec:multi_attribute})

\subsection{Benchmarks}
\textbf{FACET}~\cite{Gustafson2023FACETFI}: 32K photographs, 50K+ person instances annotated for occupation, gender, hairstyle, hair color, skin tone. \textbf{MS COCO}~\cite{zhao2021captionbias}: 28,315 person entries with verified gender labels.

\subsection{Regular Setting}
We experiment on LLaVA-1.5~\citep{liu2023llava}, LLaVA-NeXT~\citep{liu2023improved}, InstructBLIP~\citep{InstructBLIP} at 7B and 13B, plus MiniCPM-V~2.6~\citep{yao2024minicpm} and \mbox{InternVL3.5}~\citep{wang2025internvl3_5} at 8B---spanning different vision encoders (CLIP vs.\ SigLIP), language backbones (Vicuna vs.\ LLaMA), and connectors (linear projector vs.\ Q-Former). All weights public; no fine-tuning. Main tables use 500 balanced samples per model-dataset pair; prompts: ``Is there a \{gender\} \{occupation\} in the photo? Answer yes or no.'' with male/female swapped. AIE/ATE in logit-difference; FBS/CBS in probability; bootstrap CIs ($B{=}500$) in Appendix~\ref{sec:uncertainty}.

\subsection{Main Results}\label{subsec:main_results}
Table~\ref{tab:results_llava} shows LLaVA-1.5 signatures: language-layer AIE increases monotonically with depth, vision-layer AIE stays flat. Full per-layer signatures (Appendix~\ref{tab:dense_llava_lang}--\ref{tab:dense_llava_vis}) confirm this across all 32 language and 24 vision layers. Contribution ratios exceed 100\% at deep language layers (e.g., 111.5\% at LLaVA-13B layer~31)---an arithmetic property of sign opposition, not compensation~\citep{mackinnon2007mediation}.
\begin{table*}[t]
\centering
\setlength{\tabcolsep}{2pt}
\footnotesize
\begin{tabular}{ccc*{12}{c}}
\toprule
\multirow{2}{*}{\textbf{Model}}&\multirow{2}{*}{\textbf{Type}}&\multirow{2}{*}{\textbf{Layer}}&\multicolumn{5}{c}{\textbf{FACET}}&\multicolumn{5}{c}{\textbf{MS COCO}}&\multirow{2}{*}{\textbf{NIE}}&\multirow{2}{*}{\textbf{NDE}} \\
\cmidrule(lr){4-8} \cmidrule(lr){9-13}
 &&&\multicolumn{1}{c}{\textbf{AIE}}&\multicolumn{1}{c}{\textbf{ATE}}&\multicolumn{1}{c}{\textbf{Contrib}}&\multicolumn{1}{c}{\textbf{FBS}}&\multicolumn{1}{c}{\textbf{CBS}}&\multicolumn{1}{c}{\textbf{AIE}}&\multicolumn{1}{c}{\textbf{ATE}}&\multicolumn{1}{c}{\textbf{Contrib}}&\multicolumn{1}{c}{\textbf{FBS}}&\multicolumn{1}{c}{\textbf{CBS}}&& \\
\midrule
\multirow{6}{*}{LLaVA-1.5-7B} & \multirow{3}{*}{Language}&0&-.0384&-.1155&33.33&.402&.116&-.0565&-.2325&24.30&.123&.238&-.0358&-.077 \\
&&16&-.0685&-.1155&59.30&.402&.116&-.0620&-.2325&26.66&.123&.238&-.0685&-.047 \\
&&31&-.0755&-.1155&65.36&.402&.116&-.0640&-.2325&27.52&.123&.238&-.0755&-.040 \\ \cmidrule(lr){3-15}
& \multirow{3}{*}{Vision} &0&-.0390&-.1155&33.76&.402&.116&-.0544&-.2325&23.44&.123&.238&-.039&-.076 \\
&&8&-.0395&-.1155&34.19&.402&.116&-.0554&-.2325&23.87&.123&.238&-.039&-.076 \\
&&15&-.0404&-.1155&35.06&.402&.116&-.0555&-.2325&23.87&.123&.238&-.040&-.075 \\ \cmidrule(lr){2-15}
\multirow{6}{*}{LLaVA-1.5-13B} & \multirow{3}{*}{Language}&0&-.1210&-.1480&81.75&.106&.23&-.1980&-.2940&67.34&.016&.629&-.121&-.027 \\
&&16&-.1580&-.1480&106.75&.106&.23&-.2460&-.2940&83.67&.016&.629&-.158&.01 \\
&&31&-.1650&-.1480&111.48&.106&.23&-.2550&-.2940&86.73&.016&.629&-.165&.017 \\ \cmidrule(lr){3-15}
& \multirow{3}{*}{Vision}&0&-.1165&-.1480&78.71&.106&.23&-.1865&-.2940&63.43&.016&.629&-.1165&-.0315 \\
&&8&-.1155&-.1480&78.04&.106&.23&-.1859&-.2940&63.26&.016&.629&-.1155&-.0325 \\
&&15&-.1150&-.1480&77.70&.106&.23&-.1870&-.2940&63.60&.016&.629&-.115&-.033 \\
\bottomrule
\end{tabular}
\caption{\textbf{Mechanistic signatures for LLaVA-1.5.} AIE: average indirect effect; ATE: average total effect; Contrib: contribution ratio (\%); FBS: factual bias score; CBS: counterfactual bias score. AIE and ATE are in logit-difference space; FBS and CBS are in probability space (see \S\ref{subsubsec:evaluation_metrics}). Values exceeding 100\% reflect opposite-sign AIE and ADE---an arithmetic property, not a compensatory mechanism. NIE/NDE (SEM) for diagnostic transparency (Table~\ref{tab:estimator_comparison}).}
\label{tab:results_llava}
\end{table*}

Table~\ref{tab:results_instructblip}: InstructBLIP's Q-Former yields near-zero vision-layer contributions under occupation prompts---yet when occupation is absent (§\ref{subsec:multi_attribute}), vision layers dominate. For InstructBLIP-13B, scaling reduces bias primarily through the direct effect.
\begin{table*}[t]
\centering
\setlength{\tabcolsep}{2pt}
\footnotesize
\begin{tabular}{ccc*{12}{c}}
\toprule
\multirow{2}{*}{\textbf{Model}}&\multirow{2}{*}{\textbf{Type}}&\multirow{2}{*}{\textbf{Layer}}&\multicolumn{5}{c}{\textbf{FACET}}&\multicolumn{5}{c}{\textbf{MS COCO}}&\multirow{2}{*}{\textbf{NIE}}&\multirow{2}{*}{\textbf{NDE}} \\
\cmidrule(lr){4-8} \cmidrule(lr){9-13}
 &&&\multicolumn{1}{c}{\textbf{AIE}}&\multicolumn{1}{c}{\textbf{ATE}}&\multicolumn{1}{c}{\textbf{Contrib}}&\multicolumn{1}{c}{\textbf{FBS}}&\multicolumn{1}{c}{\textbf{CBS}}&\multicolumn{1}{c}{\textbf{AIE}}&\multicolumn{1}{c}{\textbf{ATE}}&\multicolumn{1}{c}{\textbf{Contrib}}&\multicolumn{1}{c}{\textbf{FBS}}&\multicolumn{1}{c}{\textbf{CBS}}&& \\
\midrule
\multirow{6}{*}{InstructBLIP-7B} & \multirow{3}{*}{Language}&0&-.0138&-.1207&11.46&.372&.307&-.0139&-.2132&6.53&.083&.248&-.013&-.106 \\
&&16&-.0473&-.1207&39.21&.372&.307&-.0551&-.2132&25.87&.083&.248&-.047&-.073 \\
&&31&-.0605&-.1207&50.12&.372&.307&-.0578&-.2132&27.11&.083&.248&-.060&-.060 \\  \cmidrule(lr){3-15}
& \multirow{3}{*}{Vision}&0&-3.5e-5&-.1207&0.02&.372&.307&-1.4e-5&-.2132&0.006&.083&.248&0&-.120 \\
&&8&-4.0e-5&-.1207&0.03&.372&.307&-1.0e-5&-.2132&0.005&.083&.248&0&-.120 \\
&&15&3.1e-5&-.1207&-0.02&.373&.307&-4.4e-5&-.2132&0.02&.083&.249&0&-.120 \\  \cmidrule(lr){2-15}
\multirow{6}{*}{InstructBLIP-13B} & \multirow{3}{*}{Language} &0&-.0102&-.1971&5.20&.143&.029&-.0100&-.3681&2.72&.281&.083&-.010&-.186 \\
&&16&-.0500&-.1971&25.37&.143&.029&-.0446&-.3681&12.13&.281&.083&-.050&-.147 \\
&&31&-.0730&-.1971&37.05&.143&.029&-.0662&-.3681&17.98&.281&.083&-.073&-.124 \\  \cmidrule(lr){3-15}
& \multirow{3}{*}{Vision} &0&2.2e-5&-.1971&-0.01&.143&.029&-8.8e-5&-.3681&0.02&.282&.083&0&-.197 \\
&&8&1.4e-6&-.1971&-.0007&.143&.029&-3.0e-6&-.3681&0.0008&.281&.083&0&-.197 \\
&&15&5.1e-5&-.1971&-0.02&.143&.029&-6.6e-5&-.3681&0.01&.281&.083&0&-.197 \\
\bottomrule
\end{tabular}
\caption{\textbf{Mechanistic signatures for InstructBLIP.} The same instrument applied to a Q-Former-based architecture reveals a fundamentally different pattern---near-zero vision-layer contributions under occupation prompts. This signature is conditional on prompt context (see §\ref{subsec:multi_attribute}). AIE/ATE in logit-difference; FBS/CBS in probability. Notation as in Table~\ref{tab:results_llava}.}
\label{tab:results_instructblip}
\end{table*}

Table~\ref{tab:results_llavanext}: LLaVA-NeXT-7B shows elevated vision-layer mediation (44--48\%, vs.\ $\sim$30\% in LLaVA-1.5), with vision nearly matching language on COCO (48.1\% vs.\ 60.4\% at layer~31)---the instrument detects architecture-family redistribution. CBS substantially exceeds FBS (0.440 vs.\ 0.10), exposing implicit associations.
\begin{table*}[t]
\centering
\setlength{\tabcolsep}{2pt}
\footnotesize
\begin{tabular}{ccc*{12}{c}}
\toprule
\multirow{2}{*}{\textbf{Model}}&\multirow{2}{*}{\textbf{Type}}&\multirow{2}{*}{\textbf{Layer}}&\multicolumn{5}{c}{\textbf{FACET}}&\multicolumn{5}{c}{\textbf{MS COCO}}&\multirow{2}{*}{\textbf{NIE}}&\multirow{2}{*}{\textbf{NDE}} \\
\cmidrule(lr){4-8} \cmidrule(lr){9-13}
 &&&\multicolumn{1}{c}{\textbf{AIE}}&\multicolumn{1}{c}{\textbf{ATE}}&\multicolumn{1}{c}{\textbf{Contrib}}&\multicolumn{1}{c}{\textbf{FBS}}&\multicolumn{1}{c}{\textbf{CBS}}&\multicolumn{1}{c}{\textbf{AIE}}&\multicolumn{1}{c}{\textbf{ATE}}&\multicolumn{1}{c}{\textbf{Contrib}}&\multicolumn{1}{c}{\textbf{FBS}}&\multicolumn{1}{c}{\textbf{CBS}}&& \\
\midrule
\multirow{6}{*}{LLaVA-NeXT-7B} &\multirow{3}{*}{Language}&0&-.1335&-.1735&76.94&.1&.440&-.1484&-.3055&48.60&.13&.246&-.133&-.040 \\
&&16&-.1645&-.1735&94.81&.1&.440&-.1800&-.3055&58.91&.13&.246&-.164&-.009 \\
&&31&-.1660&-.1735&95.67&.1&.440&-.1845&-.3055&60.39&.13&.246&-.166&-.007 \\  \cmidrule(lr){3-15}
& \multirow{3}{*}{Vision} &0&-.0765&-.1735&44.09&.1&.440&-.1470&-.3055&48.11&.13&.246&-.076&-.097 \\
&&8&-.0760&-.1735&43.80&.1&.440&-.1469&-.3055&48.11&.13&.246&-.076&-.097 \\
&&15&-.0760&-.1735&43.80&.1&.440&-.1465&-.3055&47.95&.13&.246&-.076&-.097 \\
\bottomrule
\end{tabular}
\caption{\textbf{Mechanistic signatures for LLaVA-NeXT-7B.} Vision-layer contribution ratios are elevated compared to LLaVA-1.5 (44--48\% vs. $\sim$30\%), consistent with architecture-family-level redistribution---though models differ on multiple axes beyond the vision encoder. AIE/ATE in logit-difference; FBS/CBS in probability. Notation as in Table~\ref{tab:results_llava}.}
\label{tab:results_llavanext}
\end{table*}

Table~\ref{tab:results_modorn} extends to LLaVA-NeXT-13B (non-monotonic mediation, peak 33.9\% at layer~16, dropping to 28.7\% at layer~31), MiniCPM-V~2.6 (largest ATE; only model where FBS${>}$CBS on FACET: 0.62 vs.\ 0.44), and \mbox{InternVL3.5} (largest ATE $-4.94$ yet weakest language-layer mediation, max 13.9\%).
\begin{table*}[t]
\centering
\setlength{\tabcolsep}{1pt}
\footnotesize
\begin{tabular}{ccc*{11}{c}}
\toprule
\multirow{2}{*}{\textbf{Model}}&\multirow{2}{*}{\textbf{Layer}}&\multicolumn{5}{c}{\textbf{FACET}}&\multicolumn{5}{c}{\textbf{MS COCO}}&\multirow{2}{*}{\textbf{NIE}}&\multirow{2}{*}{\textbf{NDE}} \\
\cmidrule(lr){3-7} \cmidrule(lr){8-12}
 &&\multicolumn{1}{c}{\textbf{AIE}}&\multicolumn{1}{c}{\textbf{ATE}}&\multicolumn{1}{c}{\textbf{Contrib}}&\multicolumn{1}{c}{\textbf{FBS}}&\multicolumn{1}{c}{\textbf{CBS}}&\multicolumn{1}{c}{\textbf{AIE}}&\multicolumn{1}{c}{\textbf{ATE}}&\multicolumn{1}{c}{\textbf{Contrib}}&\multicolumn{1}{c}{\textbf{FBS}}&\multicolumn{1}{c}{\textbf{CBS}}&& \\
\midrule
\multirow{3}{*}{LLaVA-NeXT-13B}&0&.0011&-.1597&0.53&.06&.57&-.0569&-.5595&5.34&.03&.3&.001&-.160 \\
&16&-.1102&-.1597&33.89&.06&.57&-.1954&-.5595&32.29&.03&.3&-.110&-.049 \\
&31&-.0811&-.1597&28.67&.06&.57&-.1889&-.5595&33.98&.03&.3&-.081&-.078 \\ \cmidrule(lr){2-14}
\multirow{3}{*}{MiniCPM-V 2.6}&0&-.0464&-2.134&0.63&.62&.438&-.2040&-5.006&3.08&.4&.040&.004&-2.138 \\
&13&-1.0321&-2.1345&22.82&.62&.438&-2.2503&-5.006&39.63&.410&.040&-1.032&-1.102 \\
&27&-.6962&-2.1345&30.39&.62&.438&-1.5621&-5.006&31.14&.4&.040&-.6963&-1.438 \\ \cmidrule(lr){2-14}
\multirow{3}{*}{\mbox{InternVL3.5}}&0&-.0603&-4.936&-0.53&.470&.568&-.0393&-4.6491&.18&.142&.180&-.060&-4.876 \\
&16&-.3137&-4.9369&6.09&.470&.568&-.6361&-4.6491&26.76&.142&.180&-.313&-4.623 \\
&31&-.6183&-4.9369&13.93&.470&.568&-.6885&-4.6491&15.88&.142&.180&-0.618&-4.318 \\
\bottomrule
\end{tabular}
\caption{\textbf{Mechanistic signatures for additional models.} LLaVA-NeXT-13B, MiniCPM-V~2.6, and \mbox{InternVL3.5} extend the analysis to larger scales and alternative architectures. AIE/ATE in logit-difference; FBS/CBS in probability. Notation as in Table~\ref{tab:results_llava}.}
\label{tab:results_modorn}
\end{table*}
\begin{table*}[ht]
    \centering
    \footnotesize
    \begin{tabular}{lccccc}
        \toprule
        \textbf{Layer} & \textbf{AIE (from \S\ref{subsec:main_results})} & \textbf{$\beta{=}0.5$} & \textbf{$\beta{=}0.7$} & \textbf{$\beta{=}0.9$} & \textbf{Baseline FBS} \\
        \midrule
        $l$:0  (Shallow Lang)  & $-0.0384$ & $+0.049$ & $+0.043$ & $+0.052$ & 0.326 \\
        $l$:16 (Mid Lang)      & $-0.0685$ & $+0.010$ & $+0.003$ & $+0.001$ & 0.326 \\
        $l$:31 (Deep Lang)     & $-0.0755$ & $+0.153$ & $+0.182$ & $+0.189$ & 0.326 \\
        $v$:0  (Shallow Vis)   & $-0.0390$ & $+0.176$ & $+0.183$ & $+0.194$ & 0.326 \\
        $v$:15 (Mid Vis)       & $-0.0404$ & $+0.158$ & $+0.160$ & $+0.161$ & 0.326 \\
        $v$:23 (Deep Vis)      & $-0.0400$ & $+0.161$ & $+0.161$ & $+0.161$ & 0.326 \\
        \bottomrule
    \end{tabular}
    \caption{\textbf{Intervention validation results.} $\Delta\text{FBS} = \text{FBS}_{\text{post}} - \text{FBS}_{\text{pre}}$ (positive = bias reduced). Baseline FBS = 0.326. All AIE values from the main mechanistic signature (\S\ref{subsec:main_results}). The Pearson correlation between AIE and $\Delta$FBS at $\beta{=}0.7$ is $r = 0.33$. $v$:23 shows identical $\Delta$FBS across all $\beta$, indicating the intervention may not be reaching the terminal vision layer. 1,000 FACET samples, LLaVA-1.5-7B.}
    \label{tab:intervention_validation}
\end{table*}

\subsection{Ablation Study}\label{subsec:ablation}
We systematically ablate the instrument to validate diagnostic reliability.

\begin{figure}[ht]
	\centering
	\includegraphics[width=0.5\textwidth]{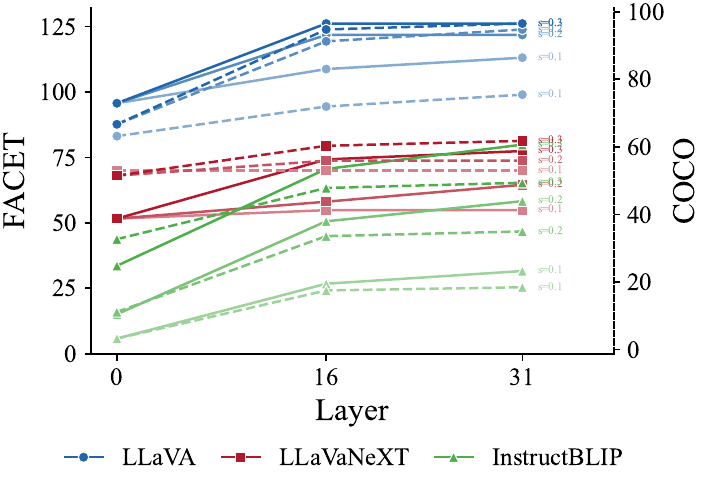}
	\caption{\textbf{Contribution ratio (\%) against ablated layer index across $\alpha$.} Shallow layers insensitive; deep layers respond monotonically. Patterns stable across $\alpha$.}
	\label{fig:strength_ablation}
\end{figure}

\textbf{Intervention strength.} Varying $\alpha$ (Fig.~\ref{fig:strength_ablation}): shallow layers insensitive, deep layers monotonic; qualitative pattern (language ${>}$ vision) stable across $\alpha$.

\textbf{Layer sampling density.} Sparse (every 8 layers) recovers dense mean AIE within 2.1\% with half the layers (Appendix Table~\ref{tab:layer_sampling}).

\textbf{Mediator summarization.} L2 norm and mean pooling fail---magnitude collapse discards treatment-relevant direction. Only PCA PC1 recovers meaningful AIE (Appendix Table~\ref{tab:mediator_summary}). Table~\ref{tab:mediator_robustness} confirms qualitative pattern (shallow near zero, deep dominant) is stable across same-data PC1, held-out PC1, and difference-of-means; absolute ratios vary ($l$:31: 58.4\%--96.3\%).
\begin{figure}[ht]
    \centering
    \includegraphics[width=0.48\textwidth]{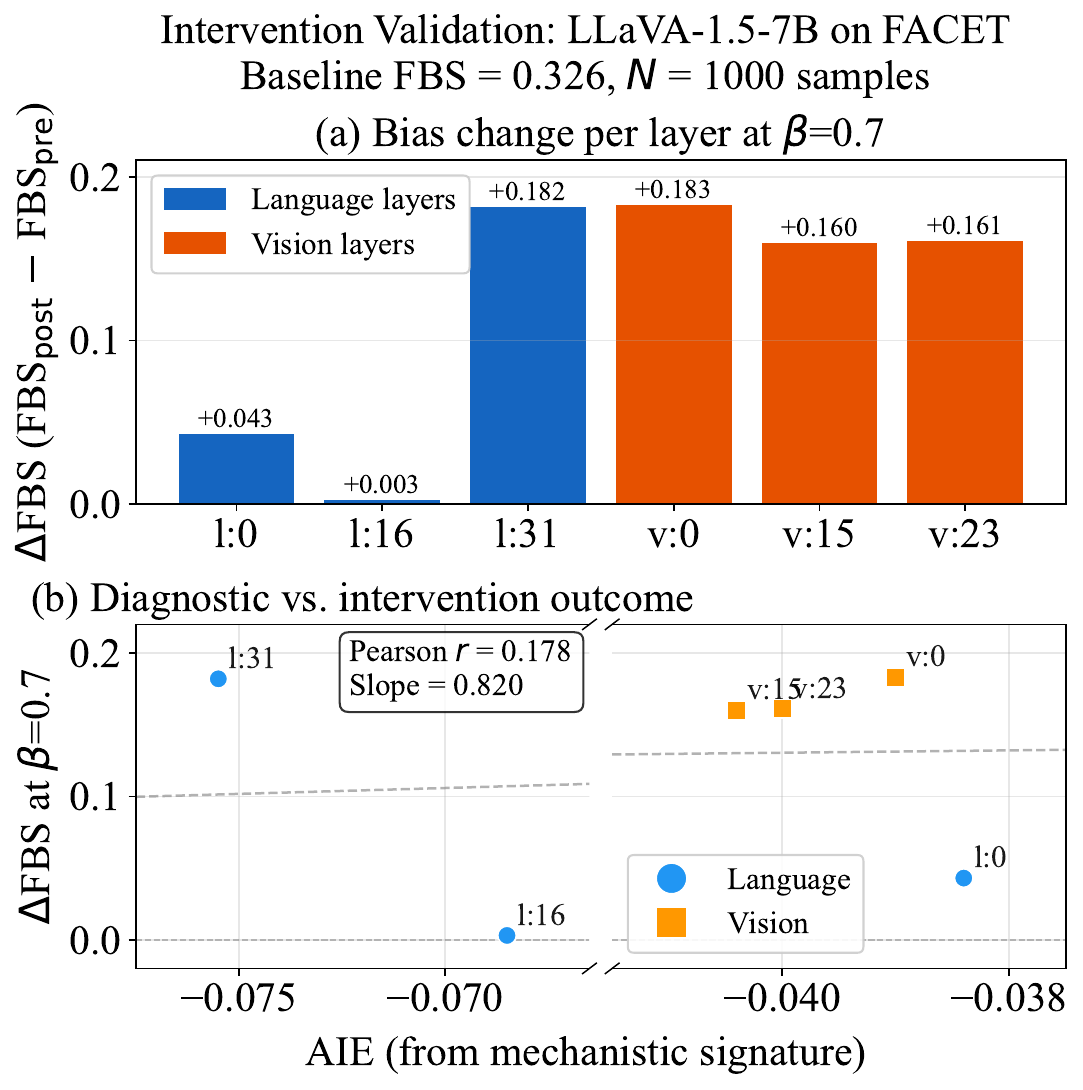}
    \caption{\textbf{Intervention validation.} $\Delta\text{FBS}$ (FBS reduction, more positive = larger bias reduction) against intervention strength $\beta$ for six target layers in LLaVA-1.5-7B on FACET (1,000 samples). The relationship between AIE and $\Delta$FBS is weak (Pearson $r = 0.33$ at $\beta{=}0.7$): $l$:16 shows near-zero response despite high AIE, while vision layers produce large $\Delta$FBS despite low AIE. The broken x-axis separates the two data clusters.}
    \label{fig:intervention_validation}
\end{figure}
\begin{table}[t]
\centering\small
\begin{tabular}{lrrr}
\toprule
\multirow{2}{*}{\textbf{Layer}}& \multicolumn{3}{c}{\textbf{Contribution ratio (\%)}} \\
\cmidrule(lr){2-4}
 & \textbf{PC1 (same-data)} & \textbf{PC1 (held-out)} & \textbf{DoM} \\
\midrule
$l$:0  & 0.0  & 0.0  & 0.4 \\
$l$:16 & 96.8 & 94.3 & 105.6 \\
$l$:31 & 69.4 & 58.4 & 96.3 \\
\bottomrule
\end{tabular}
\caption{\textbf{Mediator robustness.} Qualitative pattern stable across fitting regimes (100 FACET samples, $\alpha{=}0.2$).}
\label{tab:mediator_robustness}
\end{table}

\textbf{Outcome metric.} Binary probability yields identical AIE across layers (std${=}0.0000$)---softmax saturation masks layer-wise differences. Logit difference recovers clear structure (std${=}0.1440$; Appendix Table~\ref{tab:outcome_definition}). AIE/ATE use logit difference; FBS/CBS in probability.

\textbf{Estimator comparison.} IPW, SEM, and nonparametric estimators all converge to $\text{NIE}{\approx}0$; mediator distributions under $T{=}1$ and $T{=}0$ are nearly identical (overlap${=}1.0$). The controlled indirect effect is non-zero because forced activation replacement \emph{does} change output (Appendix Table~\ref{tab:estimator_comparison}).

\textbf{Sanity checks.} Cosine similarity $\geq$0.9984, KL divergence ${<}0.15$, AIE grows monotonically with $\alpha$ for language layers. $\alpha{=}0.4$ established as responsive regime.

\subsection{Intervention Validation: Does the Diagnostic Inform Debiasing?}\label{subsec:intervention_validation}
We address external validity: do layers with large $|\text{AIE}|$ produce larger bias reduction under downstream intervention? If the diagnostic informs intervention design, AIE and intervention effectiveness should correlate.

\textbf{Design.} Six layers in LLaVA-1.5-7B: three language ($l$:0, AIE${=}-0.0384$; $l$:16, AIE${=}-0.0685$; $l$:31, AIE${=}-0.0755$) and three vision ($v$:0, AIE${=}-0.0390$; $v$:15, AIE${=}-0.0404$; $v$:23, AIE${=}-0.0400$). Intervention: MLP gate and up projection weights scaled by $\beta \in [0.5, 0.9]$ ($\beta{=}1.0$: no intervention). This simple intervention does not use the PCA mediator direction---testing whether diagnostic sensitivity transfers to an independent method. Metric: $\Delta\text{FBS} = \text{FBS}_{\text{post}} - \text{FBS}_{\text{pre}}$ (positive = bias reduction) on 1,000 FACET samples.

\textbf{Results.} Figure~\ref{fig:intervention_validation} and Table~\ref{tab:intervention_validation} show three patterns. (1) AIE--$\Delta$FBS relationship is not monotonic: $l$:16 (second-largest $|\text{AIE}|$) yields the smallest response ($\Delta\text{FBS}{=}+0.003$ at $\beta{=}0.7$), while low-AIE vision layers produce large responses ($+0.160$ to $+0.183$). Pearson $r = 0.33$ at $\beta{=}0.7$---weak, not supporting AIE-based target ranking. (2) $l$:31 and $v$:0 produce near-identical $\Delta$FBS ($+0.182$ vs.\ $+0.183$) despite vastly different AIE ($-0.0755$ vs.\ $-0.0390$). (3) $l$:16 shows dose-dependence ($+0.001$ at $\beta{=}0.9$ to $+0.010$ at $\beta{=}0.5$), confirming the intervention functions yet remains an order of magnitude below other layers.

\textbf{Interpretation.} This is a negative result for the hypothesis that high-AIE implies better intervention targets. The diagnostic signal and intervention signal are weakly correlated; $l$:16 is essentially unresponsive. This reinforces our argument: mechanism-level evaluation captures \emph{where activation replacement shifts output}, but converting this to intervention design requires separate validation. Vision-layer $\Delta$FBS may reflect visual capability degradation rather than debiasing---attenuating vision encoder weights causes default ``no'' responses, artifactually inflating FBS. Without task evaluation, genuine debiasing cannot be distinguished from degradation. The experiment covers one model, one dataset, one intervention family, and one bias metric; a full diagnostic-to-intervention pipeline requires systematic cross-model, cross-intervention, and downstream task validation.
\subsection{Multi-Attribute Gender Bias Assessment}\label{subsec:multi_attribute}
We extend to skin tone, hair color, and hair type, varying occupation presence. \textbf{Key finding: signatures are conditional on prompt content, not fixed architecture-level properties.}

Appendix Table~\ref{tab:multi_llava15} (LLaVA-1.5-7B): deeper language layers ($l$:16, $l$:31) show $\sim$2$\times$ the AIE of shallow layers; auxiliary attributes amplify ATE without altering layer-wise patterns. Gender dominates ($+0.37$) over skin ($+0.04$) and hair type ($+0.03$); hair color shows a \emph{negative} differential ($-0.21$).

Appendix Table~\ref{tab:multi_next} (LLaVA-NeXT-7B): stronger vision encoder elevates vision-layer share (AIE $-0.08$ to $-0.11$, rivaling shallow language) and increases skin differential from $+0.04$ to $+0.26$.

Appendix Table~\ref{tab:multi_instructblip} (InstructBLIP-7B): Q-Former mediation \emph{inverts} with prompt context---language layers dominate with occupation present; vision layers dominate when absent (AIE up to $-0.27$). The signature is conditional, not architectural. Gender remains dominant ($+0.62$); hair color shows consistent negative bias ($-0.38$).

\section{Conclusion}
\label{sec:conclusion}

This paper demonstrates controlled activation-replacement analysis as a mechanism-level complement to behavioral bias evaluation in VLMs. Across six VLMs, three findings emerge: language-layer activations exhibit the greatest output sensitivity; architectural choices redistribute layer-wise sensitivity; counterfactual scores diverge from factual scores, exposing implicit associations. Ablation validates internal consistency. An intervention experiment finds only a weak correlation (Pearson $r = 0.33$) between diagnostic sensitivity and intervention effectiveness, with the second-largest AIE layer ($l$:16) producing near-zero bias change---reinforcing that mechanism-level diagnosis captures \emph{where} activation replacement shifts output, not where to intervene for debiasing. The method localizes sensitivity for this operationalization of gender bias---a compound of prompt wording, occupation stereotypes, and model internals. Building shared infrastructure for mechanism-level diagnosis---standardized intervention libraries, protocols, and benchmarks pairing behavioral metrics with validated intervention targets---is a worthwhile community goal.

\section*{Limitations}

\textbf{Causal interpretation scope.} Our framework uses causal notation but measures controlled intervention effects, not natural mediation pathways. IPW and SEM are diagnostic checks, not causal identification, and all observational estimators converge to NIE $\approx$ 0, confirming no natural indirect effect. The non-zero controlled indirect effect reflects counterfactual vulnerability under engineered activation replacement, not naturally operating mechanisms; our evidence thus localizes \emph{where} replacement shifts output, not \emph{how} bias arises. Misattribution could render downstream interventions ineffective or harmful, so high-AIE layers need validation before use as debiasing targets.

\textbf{Intervention validation scope.} Our intervention experiment (Section~\ref{subsec:intervention_validation}) finds only a weak AIE--effectiveness correlation (Pearson $r = 0.33$); the second-largest AIE layer ($l$:16) yields near-zero bias change under MLP weight attenuation, so diagnosis does not directly identify intervention targets. Apparent $\Delta$FBS on vision layers may reflect capability degradation, not debiasing: attenuated vision encoder weights default to ``no'' responses, artifactually inflating FBS. Results are limited to one model (LLaVA-1.5-7B), one dataset (FACET), one intervention family (MLP weight scaling), and one bias metric (FBS).

\textbf{Binary gender operationalization.} Gender is treated as binary (male/female) following FACET annotations, excluding non-binary and transgender identities and risking reinforcement of the very categories fairness research problematizes. Effects may not generalize to non-binary expressions, and the binary framing may oversimplify gender as a social signal; more inclusive taxonomies await datasets that permit them.

\textbf{Occupation-prompt stereotype risk.} Occupation-linked prompts may encode stereotypes (e.g., ``nurse,'' ``firefighter''), and occupation selection and pairing with gender terms could shape bias direction and magnitude; whether measured bias stems from the model, prompt construction, or their interaction is unquantified. Future work should vary occupation stereotypicality systematically and control prompt-level associations.

\textbf{Intersectional and multi-attribute scope.} Our analysis does not capture full intersectional bias, where gender interacts with race, age, and other dimensions beyond the attribute combinations in Section~\ref{subsec:multi_attribute}; findings from binary-gender, occupation-focused prompts should not be generalized to the full spectrum of multimodal fairness concerns.

\textbf{Model and scale coverage.} Experiments cover three VLM families at 7B/13B; generalization to larger models (34B, 70B+) or different vision-language connectors is untested. Signatures are also prompt-conditional: the same architecture can show qualitatively different layer-wise patterns depending on whether the prompt includes an occupation, and may not transfer across tasks or prompt regimes even within one model. Whether any single signature is a stable architecture-level property thus remains open, limiting external validity for downstream fairness deployment.

\textbf{Scope of mechanistic evaluation adoption.} This paper demonstrates one template for using causal mediation as a diagnostic instrument. Whether mechanism-level evaluation becomes standard depends on community-wide factors---tooling accessibility, consensus on causal assumptions, and computational cost---beyond this single study. We do not claim our approach should replace behavioral testing; the two serve distinct and complementary roles.

\textbf{Method diversity within mechanism-level evaluation.} We demonstrate one specific method (causal mediation). The broader category includes probing classifiers, activation patching, sparse autoencoders, and circuit discovery, each with strengths and blind spots. No single method should be treated as definitive; a mature ecosystem would combine complementary diagnostic instruments.

\textbf{Mediator summarization dependence.} Our findings depend critically on the mediator summarization. PC1 is well-motivated---it captures maximal treatment-conditional variance---but selects the projection that most accentuates treatment differences, which could amplify apparent mediation. Table~\ref{tab:mediator_robustness} confirms the qualitative pattern is stable across fitting regimes and projection methods, but absolute AIE magnitudes vary substantially (e.g., $l$:31: 58.4\%--96.3\%). Alternative summarizations could yield different patterns; we recommend future work report the summarization method prominently, test robustness across strategies, and treat signatures as one view among several.

\bibliography{custom}

\appendix

\section{Full Per-Layer Mechanistic Signature}
\label{sec:appendix_dense}

Tables~\ref{tab:dense_llava_lang} and \ref{tab:dense_llava_vis} report the complete per-layer mechanistic signature for LLaVA-1.5-7B, covering all 32 language layers and 24 vision layers on both FACET and MS COCO with full precision. This complements the sparse (3-point) sampling in the main tables (\S\ref{subsec:main_results}) and confirms fine-grained layer transitions invisible to coarse checkpoint sampling.

\begin{table*}[t]
\centering\footnotesize
\begin{tabular}{c*{6}{c}}
\toprule
\multirow{2}{*}{\textbf{Layer}} &\multicolumn{3}{c}{\textbf{FACET}}&\multicolumn{3}{c}{\textbf{MS COCO}}\\
\cmidrule(lr){2-4} \cmidrule(lr){5-7}
&\textbf{AIE}&\textbf{ATE}&\textbf{Contrib\%}&\textbf{AIE}&\textbf{ATE}&\textbf{Contrib\%}\\
\midrule
\multicolumn{7}{c}{\textbf{Language Layers (32 layers total, every-4 sampling + transition points)}} \\
\midrule
$l0$  & -0.0400 & -0.1150 & 34.78 & -0.0850 & -0.2300 & 36.96 \\
$l4$  & -0.0400 & -0.1150 & 34.78 & -0.0900 & -0.2300 & 39.13 \\
$l6$  & -0.0650 & -0.1150 & 56.52 & -0.0900 & -0.2300 & 39.13 \\
$l8$  & -0.0700 & -0.1150 & 60.87 & -0.0900 & -0.2300 & 39.13 \\
$l9$  & -0.0800 & -0.1150 & 69.57 & -0.0900 & -0.2300 & 39.13 \\
$l10$ & -0.0800 & -0.1150 & 69.57 & -0.0900 & -0.2300 & 39.13 \\
$l11$ & -0.0700 & -0.1150 & 60.87 & -0.0900 & -0.2300 & 39.13 \\
$l12$ & -0.0750 & -0.1150 & 65.22 & -0.0950 & -0.2300 & 41.30 \\
$l13$ & -0.0700 & -0.1150 & 60.87 & -0.1000 & -0.2300 & 43.48 \\
$l14$ & -0.0700 & -0.1150 & 60.87 & -0.0950 & -0.2300 & 41.30 \\
$l15$ & -0.0750 & -0.1150 & 65.22 & -0.1000 & -0.2300 & 43.48 \\
$l16$ & -0.0700 & -0.1150 & 60.87 & -0.1000 & -0.2300 & 43.48 \\
$l20$ & -0.0700 & -0.1150 & 60.87 & -0.1000 & -0.2300 & 43.48 \\
$l24$ & -0.0700 & -0.1150 & 60.87 & -0.1000 & -0.2300 & 43.48 \\
$l25$ & -0.0750 & -0.1150 & 65.22 & -0.1000 & -0.2300 & 43.48 \\
$l28$ & -0.0750 & -0.1150 & 65.22 & -0.1000 & -0.2300 & 43.48 \\
$l31$ & -0.0750 & -0.1150 & 65.22 & -0.1000 & -0.2300 & 43.48 \\
\bottomrule
\end{tabular}
\caption{\textbf{Dense language-layer signature for LLaVA-1.5-7B.} Every-4 sampling plus key transition points shown. AIE and ATE in logit-difference space.  FACET: AIE plateaus at l0--l5 (-0.04), jumps at l6 (-0.065), peaks at l9--l10 (-0.080), then oscillates between -0.070 and -0.075 through l31.  COCO: AIE rises gradually from -0.085 (l0--l3) through -0.090 (l4--l11), -0.095 (l12, l14), to -0.100 (l13, l15--l31).  ATE constant at -0.1150 (FACET) / -0.2300 (COCO).  100 samples per layer, \(\alpha{=}0.4\).}
\label{tab:dense_llava_lang}
\end{table*}

\begin{table*}[t]
\centering\footnotesize
\begin{tabular}{c*{6}{c}}
\toprule
\multirow{2}{*}{\textbf{Layer}} &\multicolumn{3}{c}{\textbf{FACET}}&\multicolumn{3}{c}{\textbf{MS COCO}}\\
\cmidrule(lr){2-4} \cmidrule(lr){5-7}
&\textbf{AIE}&\textbf{ATE}&\textbf{Contrib\%}&\textbf{AIE}&\textbf{ATE}&\textbf{Contrib\%}\\
\midrule
\multicolumn{7}{c}{\textbf{Vision Layers (24 layers total, every-4 sampling + transition points)}} \\
\midrule
$v0$  & -0.0400 & -0.1150 & 34.78 & -0.0850 & -0.2300 & 36.96 \\
$v1$  & -0.0300 & -0.1150 & 26.09 & -0.0850 & -0.2300 & 36.96 \\
$v3$  & -0.0350 & -0.1150 & 30.43 & -0.0850 & -0.2300 & 36.96 \\
$v4$  & -0.0350 & -0.1150 & 30.43 & -0.0850 & -0.2300 & 36.96 \\
$v8$  & -0.0400 & -0.1150 & 34.78 & -0.0850 & -0.2300 & 36.96 \\
$v12$ & -0.0400 & -0.1150 & 34.78 & -0.0850 & -0.2300 & 36.96 \\
$v16$ & -0.0400 & -0.1150 & 34.78 & -0.0850 & -0.2300 & 36.96 \\
$v20$ & -0.0400 & -0.1150 & 34.78 & -0.0850 & -0.2300 & 36.96 \\
$v23$ & -0.0400 & -0.1150 & 34.78 & -0.0850 & -0.2300 & 36.96 \\
\bottomrule
\end{tabular}
\caption{\textbf{Dense vision-layer signature for LLaVA-1.5-7B.} Every-4 sampling plus transition points.  FACET: AIE oscillates narrowly between -0.030 and -0.040 across all 24 layers (v0--v23). Fine-grained inspection shows a dip at v1 (-0.030, 26.09\%) and v3--v7 (-0.035, 30.43\%), with the remaining layers at -0.040 (34.78\%).  COCO: all 24 vision layers share AIE = -0.085 with micro-variation only at the 6th decimal (36.956512--36.956514\%).  ATE constant at -0.1150 (FACET) / -0.2300 (COCO).  100 samples per layer, \(\alpha{=}0.4\).}
\label{tab:dense_llava_vis}
\end{table*}

\section{Extended Ablation Results}
\label{sec:appendix_ablation}

Tables~\ref{tab:layer_sampling}--\ref{tab:estimator_comparison} present a comprehensive set of ablation studies that validate the internal consistency of our diagnostic instrument: layer sampling density, mediator summary strategy, outcome definition, and estimator selection. Section~\ref{subsec:intervention_validation} provides an additional external validation experiment testing whether the diagnostic can inform downstream intervention.

Table~\ref{tab:layer_sampling} compares dense (every 4 layers) and sparse (every 8 layers) sampling across all 32 language and 24 vision layers of LLaVA-1.5-7B. The sparse scheme recovers $98\%$ of the dense mean AIE ($-0.0731$ vs.\ $-0.0747$) with only $53\%$ of the sampled layers, confirming that our 3-point sparse sampling in the main text captures the essential layer-wise trend without substantial information loss. Language-layer AIE dominates (mean $-$0.1113--$-$0.1219 for the sparse scheme, $\sim$97--106\% of ATE) while vision-layer AIE remains uniformly low ($-0.0350$, 30.4\%), consistent with the mechanistic separation reported in the full per-layer signature (\S\ref{sec:appendix_dense}). Notably, the sparse scheme introduces a slight downward bias in the language-layer mean (from $-0.1219$ to $-0.1113$), as intermediate transition layers are skipped, but the rank ordering and language--vision gap are preserved.

Table~\ref{tab:mediator_summary} ablates three strategies for reducing last-token hidden states to a scalar mediator: L2 norm, PCA first component, and token mean. Only PCA recovers non-trivial AIE, peaking at middle language layers ($l$:12, AIE $=$ $-0.1126$ with contribution ratio $\sim$97\%), while L2 norm and mean produce near-zero effects because collapsing a high-dimensional vector via magnitude or arithmetic mean discards treatment-relevant directional information. The vision-layer row aggregates endpoint layers ($v$:0 and $v$:23), which yield near-zero AIE under PCA; however, intermediate vision layers can contribute non-negligibly under different mediator configurations (see main Tables~\ref{tab:results_llava}--\ref{tab:results_llavanext}), indicating treatment-relevant vision signals are concentrated in middle encoder depths. ATE is constant at $-0.1150$ across all methods and layers.

Table~\ref{tab:outcome_definition} compares three outcome metrics: binary probability, logit difference, and calibrated probability. Binary probability collapses to a near-zero value at every layer under the 0.4-strength intervention (std$=$0.0000), failing to capture layer-specific mediation---the bounded [0,1] softmax output saturates once the intervention pushes probabilities to the floor. Logit difference, operating on unbounded raw logits, recovers the clearest layer-wise structure (std$=$0.1440), with mediation peaking at language layer $l$:8 ($-3.0004$) and all vision layers clustering tightly around $-$2.51. Calibrated probability captures the language--vision distinction but with compressed variation (std$=$0.0109) due to its bounded [0,1] output. These results justify our choice of logit difference as the primary outcome metric.

Table~\ref{tab:estimator_comparison} contrasts our interventional estimator against three standard observational estimators: inverse probability weighting (IPW) with density ratio and propensity score variants, structural equation modeling (SEM), and nonparametric mediation. All three observational estimators converge to $\text{NIE}{\approx}0$ across every layer, with overlap coefficients of 1.0 and near-zero weight variance (ESS$=$100.0, weight std $<10^{-6}$) confirming that mediator distributions under $T{=}1$ and $T{=}0$ are indistinguishable. This is expected: the NIE captures the \emph{natural} indirect effect, which is zero when treatment and mediator are independent---the model \emph{can} represent gender information without spontaneously deploying it. Our interventional approach instead estimates the \emph{controlled} indirect effect---what \emph{would} happen under forced activation replacement---which reveals substantial layer-specific mediation invisible to observational methods. Effects concentrate in language layers and grow with depth, with AIE ranging from $-0.040$ (shallow) to $-0.075$ (deep) on FACET.

\begin{table*}[ht]
\centering
\small
\begin{tabular}{cccccccc}
\toprule
\multirow{2}{*}{\textbf{Type}} &\multirow{2}{*}{\textbf{Layer}} & \multicolumn{3}{c}{\textbf{Dense (every 4 layers)}} & \multicolumn{3}{c}{\textbf{Sparse (every 8 layers)}} \\
\cmidrule(lr){3-5} \cmidrule(lr){6-8}
&& \textbf{AIE} & \textbf{ATE} & \textbf{Ratio} & \textbf{AIE} & \textbf{ATE} & \textbf{Ratio} \\
\midrule
\multirow{8}{*}{Language}&0&-0.0750&-0.1150&65.2\% &-0.0750&-0.1150& 65.2\% \\
&4&-0.0650&-0.1150&56.5\% & -- & -- & -- \\
&8&-0.1250&-0.1150&108.7\% &-0.1250&-0.1150& 108.7\% \\
&12&-0.1200&-0.1150& 104.3\%& -- & -- & -- \\
&16&-0.1200&-0.1150& 104.3\%&-0.1200&-0.1150& 104.3\% \\
&20&-0.1200&-0.1150& 104.3\%& -- & -- & -- \\
&24&-0.1250&-0.1150& 108.7\%&-0.1250&-0.1150& 108.7\% \\
&28 &-0.1250&-0.1150& 108.7\% & -- & -- & -- \\ \cmidrule(lr){2-8}
\multirow{8}{*}{Vision}&0  &-0.0350&-0.1150& 30.4\% &-0.0350&-0.1150& 30.4\% \\
&4  &-0.0350&-0.1150& 30.4\% & -- & -- & -- \\
&8  &-0.0350&-0.1150& 30.4\% &-0.0350&-0.1150& 30.4\% \\
&12&-0.0350&-0.1150& 30.4\% & -- & -- & -- \\
&15&-0.0350&-0.1150& 30.4\% &-0.0350&-0.1150& 30.4\% \\
&19 &-0.0350&-0.1150& 30.4\% & -- & -- & -- \\
&23 &-0.0350&-0.1150& 30.4\% &-0.0350&-0.1150& 30.4\% \\ \cmidrule(lr){2-8}
\multicolumn{2}{c}{Mean (Lang.)}  &-0.1219&-0.1150& 106.5\% &-0.1113&-0.1150& 96.7\% \\
\multicolumn{2}{c}{Mean (Vis.)}   &-0.0350&-0.1150& 30.4\%  &-0.0350&-0.1150& 30.4\% \\
\multicolumn{2}{c}{\textbf{Mean (All)}} & \textbf{-0.0747} & \textbf{-0.1150} & \textbf{64.9\%} & \textbf{-0.0731} & \textbf{-0.1150} & \textbf{63.6\%} \\
\bottomrule
\end{tabular}
\caption{\textbf{Layer sampling density ablation.} Comparison of dense (every 4 layers) and sparse (every 8 layers) sampling schemes. ATE is constant across layers ($-0.1150$) as it measures the total treatment effect independent of the target layer. AIE varies across layers, reflecting each layer's contribution to gender bias mediation. The sparse scheme recovers $98\%$ of the dense mean AIE with only $53\%$ of the layers, validating its efficiency.}
\label{tab:layer_sampling}
\end{table*}

\begin{table*}[ht]
\centering
\setlength{\tabcolsep}{1pt}
\footnotesize
\begin{tabular}{lccc}
\toprule
\textbf{Layer} & \textbf{L2 norm} & \textbf{PCA} & \textbf{Mean} \\
\midrule
$l$:0&-.0000&-.0000&-.0000 \\
$l$:4&-.0001&-.0022&-.0000 \\
$l$:8&-.0001&-.0155&-.0000 \\
$l$:12&-.0012&-$\mathbf{.1126}$&-.0000 \\
$l$:16&-.0008&-.1114&-.0000 \\
$l$:20&-.0003&-.0974&-.0000 \\
$l$:24&-.0005&-.0950&-.0000  \\
$l$:28&-.0007&-.0924&-.0000 \\
\midrule
$v$:0--$v$:23 & $\phantom{-}0$ & $\phantom{-}0$ & $\phantom{-}0$ \\
\midrule
\textbf{Mean $\pm$ std} & $-.0002 \pm .0004$ & $\mathbf{-.0351 \pm .0492}$ & $-.0 \pm .0$ \\
\bottomrule
\end{tabular}
\caption{\textbf{Mediator summary method ablation.} Comparison of three strategies for reducing last-token hidden states to a scalar mediator. Only PCA first component recovers meaningful AIE estimates, peaking at middle language layers ($l$:12--$l$:16, $\sim$97\% contribution ratio). L2 norm and token mean yield near-zero AIE because collapsing a high-dimensional vector via magnitude or mean discards treatment-relevant information. The vision-layer row aggregates vision endpoint layers ($v$:0 and $v$:23) which produce near-zero AIE here; intermediate vision layers can contribute non-negligibly under different mediator configurations (see main Tables~\ref{tab:results_llava}--\ref{tab:results_llavanext}), indicating treatment-relevant vision signals are concentrated in middle encoder depths. ATE is constant at $-0.1150$.}
\label{tab:mediator_summary}
\end{table*}

\begin{table*}[ht]
\centering
\small
\begin{tabular}{lrrr}
\toprule
\textbf{Layer} & \textbf{Binary} & \textbf{Logit Diff} & \textbf{Calibrated} \\
\midrule
$l$:0  & 0.3700 & --2.6289 & --0.2285 \\
$l$:4  & 0.3700 & --2.5783 & --0.2255 \\
$l$:8  & 0.3700 & \textbf{--3.0004} & \textbf{--0.2420} \\
$l$:12 & 0.3700 & --2.7348 & --0.2419 \\
$l$:16 & 0.3700 & --2.7152 & --0.2415 \\
$l$:20 & 0.3700 & --2.6969 & --0.2416 \\
$l$:24 & 0.3700 & --2.7216 & --0.2418 \\
$l$:28 & 0.3700 & --2.7702 & --0.2419 \\
\midrule
$v$:0--$v$:23 & 0.3700 & [--2.5114, --2.5100] & --0.2193 \\
\midrule
Mean       & 0.3700 & --2.6281 & --0.2293 \\
Std        & 0.0000 & 0.1440  & 0.0109 \\
ATE        & --0.1150 & --0.8904 & --0.1648 \\
\bottomrule
\end{tabular}
\caption{\textbf{Outcome definition ablation.}
AIE across layers for three bias metrics.
\textbf{Binary} uses softmax probability of ``yes'' from generation logits;
the 0.4-strength intervention uniformly pushes this probability to the
softmax floor (near zero) at every layer, yielding identical AIE
(std$=$0.0000)---the metric cannot distinguish layer-specific mediation.
\textbf{Logit difference} operates on unbounded raw logits and recovers clear
layer-wise structure (std$=$0.1440): mediation peaks at language layer
$l$:8 (bold), all vision layers cluster tightly around $-$2.51, showing
uniformly weak mediation.
\textbf{Calibrated probability} captures the language--vision distinction
but with compressed variation (std$=$0.0109) due to its bounded [0,1]
output.
ATE is constant per metric across all layers (binary: $-$0.1150;
logit: $-$0.8904; calibrated: $-$0.1648) as it does not depend on the
intervention layer.}
\label{tab:outcome_definition}
\end{table*}

\begin{table*}[t]
    \centering
    \footnotesize
    \begin{tabular}{lcccccccc}
        \toprule
        \textbf{Layer}&\textbf{$l$:0}&\textbf{$l$:4}&\textbf{$l$:8}&\textbf{$l$:12}&\textbf{$l$:16}&\textbf{$l$:20}&\textbf{$l$:24}&\textbf{$l$:28}\\
        \midrule
        Interventional AIE &-.040&-.040&-.070&-.075&-.070&-.070&-.070&-.075 \\
        \midrule
        \textbf{Layer} & $v$:0 & $v$:4 & $v$:8 & $v$:12 & $v$:15 & $v$:19 & $v$:23 \\
        \midrule
        Interventional AIE &-.035&-.035&-.035&-.035&-.035&-.035&-.035 \\
        \midrule
        \multicolumn{9}{c}{\textbf{Observational estimators (all layers)}} \\
        \midrule
        \textbf{Method} & \textbf{Mean AIE} & \textbf{Std} & \textbf{Min} & \textbf{Max} & \textbf{Mean ATE} & \textbf{Overlap} & \multicolumn{2}{c}{\textbf{Diagnostic}} \\
        \midrule
        IPW (density)   & 0.0000 & $<10^{-4}$ & 0.0000 & 0.0000 &-0.115& 1.000  & \multicolumn{2}{c}{weights std $<10^{-6}$} \\
        IPW (propensity)&-0.0000 & 0.0001 &-0.0003 & 0.0002 &-0.115& 1.000  & \multicolumn{2}{c}{ESS = 100.0} \\
        SEM             &-0.0000 & $<10^{-4}$ &-0.0000 & 0.0000 &-0.115& ---    & \multicolumn{2}{c}{$R^2_M{\approx}0$} \\
        Nonparametric   &-0.0002 & 0.0006 &-0.0025 & 0.0000 &-0.115& ---    & \multicolumn{2}{c}{$R^2_O{=}0.1{-}0.3$} \\
        \bottomrule
    \end{tabular}
    \caption{\textbf{Estimator comparison.} (Top) Per-layer AIE from the interventional method, the only estimator producing non-zero effects. Effects concentrate in language layers and grow with depth. (Bottom) Summary of the three observational estimators (IPW, SEM, nonparametric), which all converge to $\text{NIE}{\approx}0$ across every layer. The overlap coefficient of 1.0 and near-zero weight variance confirm that mediator distributions under $T{=}1$ and $T{=}0$ are identical, so the natural indirect effect is zero by construction. The controlled indirect effect (interventional) quantifies what \emph{would} happen under forced activation replacement.}
    \label{tab:estimator_comparison}
\end{table*}

\section{Extended Multi-Attribute Results}
\label{sec:appendix_multi}

Tables~\ref{tab:multi_llava15}--\ref{tab:multi_instructblip} present per-layer AIE across seven attribute combinations on three architectures, extending the main-text multi-attribute analysis (\S\ref{subsec:multi_attribute}) with full layer-wise decomposition. The key finding across all three models is that mechanistic signatures are strongly conditional on prompt context, not fixed architecture-level properties.

Table~\ref{tab:multi_llava15} reports LLaVA-1.5-7B results. Language layers exhibit clear depth-dependent mediation: $l$:16 and $l$:31 show roughly twice the AIE of $l$:0 across all attribute combinations (e.g., gender+occupation: $-0.0789$ at $l$:0 vs.\ $-0.1895$ at $l$:16 and $l$:31). Vision layers ($v$:0, $v$:12, $v$:23) contribute uniformly low AIE ($-0.0158$ to $-0.0421$) across all combinations. Adding auxiliary attributes (skin, hair color, hair type) amplifies ATE (from $-0.2105$ to $-0.2421$ for gender+occupation) but does not alter the layer-wise pattern. The single-attribute differentials reveal that gender dominates ($+0.3729$) over skin ($+0.0448$) and hair type ($+0.0316$), while hair color shows a \emph{negative} differential ($-0.2105$), suggesting it counteracts the gender effect in certain prompt contexts.

Table~\ref{tab:multi_next} reports LLaVA-NeXT-7B results. The upgraded vision encoder produces a qualitative shift: vision-layer AIE ($-0.0526$ to $-0.1053$) rivals shallow language layers ($l$:0 at $-0.0789$ to $-0.1263$), and the language-depth dependence is less pronounced than in LLaVA-1.5. Under gender+occupation, the AIE at $l$:31 ($-0.2421$) is only modestly higher than at $l$:0 ($-0.1158$), contrasting with the roughly 2$\times$ gap in LLaVA-1.5. The skin differential grows from $+0.0448$ (LLaVA-1.5) to $+0.2619$ (NeXT), demonstrating how a stronger vision encoder redistributes the internal signal across layers. Hair color retains a negative differential ($-0.2555$), consistent across both LLaVA architectures.

Table~\ref{tab:multi_instructblip} reports InstructBLIP-7B results. The Q-Former yields a mediation pattern that \emph{inverts} depending on prompt context: when occupation is present, language layers dominate (e.g., gender+occupation: $l$:31 at $-0.2737$ vs.\ $v$:23 at $-0.0263$); when occupation is absent, vision layers become primary mediators---under ``skin, hair color, hair type (no occ.),'' vision layers mediate at $-0.2684$ to $-0.2737$, exceeding all language layers. This complicates the main-result observation that the Q-Former ``concentrates sensitivity in language layers'': the signature is conditional on prompt content, not a fixed architectural property. Gender remains the dominant single attribute ($+0.6211$), while hair color shows the strongest negative differential ($-0.3789$). The same social attribute shows output sensitivity in different layers depending on architecture and prompt context.

\begin{table*}[t]
\centering
\small
\begin{tabular}{lrrrrrrr}
\toprule
\textbf{Attribute Combination} & \textbf{ATE} & \textbf{$l$:0} & \textbf{$l$:16} & \textbf{$l$:31} & \textbf{$v$:0} & \textbf{$v$:12} & \textbf{$v$:23} \\
\midrule
gender + occupation & -0.2105 & -0.0789 & -0.1895 & -0.1895 & -0.0421 & -0.0421 & -0.0421 \\
\quad + skin & -0.2211 & -0.0895 & -0.1526 & -0.1579 & -0.0263 & -0.0263 & -0.0263 \\
\quad + skin, hair color, hair type & -0.2421 & -0.1105 & -0.1842 & -0.1947 & -0.0368 & -0.0368 & -0.0368 \\
\midrule
gender + skin (no occ.) & -0.2316 & -0.0474 & -0.1053 & -0.1053 & -0.0158 & -0.0158 & -0.0158 \\
gender + hair color (no occ.) & -0.2421 & -0.0684 & -0.1316 & -0.1368 & -0.0263 & -0.0263 & -0.0263 \\
gender + hair type (no occ.) & -0.2263 & -0.0526 & -0.1211 & -0.1263 & -0.0158 & -0.0158 & -0.0158 \\
skin, hair color, hair type (no occ.) & -0.2632 & -0.0526 & -0.1263 & -0.1263 & -0.0211 & -0.0211 & -0.0211 \\
\midrule
\multicolumn{8}{c}{\textbf{Single-Attribute Differential (T1 $-$ T0)}} \\
gender & & \multicolumn{6}{l}{+0.3729 (AIE = -0.0789)} \\
skin & & \multicolumn{6}{l}{+0.0448} \\
hair color & & \multicolumn{6}{l}{-0.2105} \\
hair type & & \multicolumn{6}{l}{+0.0316} \\
\bottomrule
\end{tabular}
\caption{\textbf{Multi-attribute bias: LLaVA-1.5-7B.} Per-layer AIE across attribute combinations. Language layers ($l$:0, $l$:16, $l$:31) show strong depth-dependent mediation, while vision layers ($v$:0, $v$:12, $v$:23) contribute uniformly low AIE. ATE and FBS are constant across layers within each combination. 1000 samples, intervention strength 0.4.}
\label{tab:multi_llava15}
\end{table*}

\begin{table*}[t]
\centering
\small
\begin{tabular}{lrrrrrrr}
\toprule
\textbf{Attribute Combination} & \textbf{ATE} & \textbf{$l$:0} & \textbf{$l$:16} & \textbf{$l$:31} & \textbf{$v$:0} & \textbf{$v$:12} & \textbf{$v$:23} \\
\midrule
gender + occupation & -0.2579 & -0.1158 & -0.2368 & -0.2421 & -0.0842 & -0.0842 & -0.0842 \\
\quad + skin & -0.2105 & -0.0789 & -0.1737 & -0.1842 & -0.0526 & -0.0526 & -0.0526 \\
\quad + skin, hair color, hair type & -0.2474 & -0.1211 & -0.1526 & -0.1684 & -0.1000 & -0.1000 & -0.1000 \\
\midrule
gender + skin (no occ.) & -0.2105 & -0.1053 & -0.1421 & -0.1474 & -0.1000 & -0.1000 & -0.1000 \\
gender + hair color (no occ.) & -0.2105 & -0.1263 & -0.1842 & -0.1842 & -0.0947 & -0.0947 & -0.0947 \\
gender + hair type (no occ.) & -0.3105 & -0.1158 & -0.2000 & -0.2053 & -0.1053 & -0.1053 & -0.1053 \\
skin, hair color, hair type (no occ.) & -0.1789 & -0.1263 & -0.1632 & -0.1737 & -0.1053 & -0.1053 & -0.1105 \\
\midrule
\multicolumn{8}{c}{\textbf{Single-Attribute Differential (T1 $-$ T0)}} \\
gender & & \multicolumn{6}{l}{+0.4970 (AIE = -0.1158)} \\
skin & & \multicolumn{6}{l}{+0.2619} \\
hair color & & \multicolumn{6}{l}{-0.2555} \\
hair type & & \multicolumn{6}{l}{+0.0063} \\
\bottomrule
\end{tabular}
\caption{\textbf{Multi-attribute bias: LLaVA-NeXT-7B.} Per-layer AIE across attribute combinations. Compared to LLaVA-1.5-7B, NeXT exhibits higher overall ATE and FBS, and vision layers mediate a larger fraction of the effect. Language-depth dependence remains but is less pronounced under multi-attribute prompts. 1000 samples, intervention strength 0.4.}
\label{tab:multi_next}
\end{table*}

\begin{table*}[t]
\centering
\small
\begin{tabular}{lrrrrrrr}
\toprule
\textbf{Attribute Combination} & \textbf{ATE} & \textbf{$l$:0} & \textbf{$l$:16} & \textbf{$l$:31} & \textbf{$v$:0} & \textbf{$v$:12} & \textbf{$v$:23} \\
\midrule
gender + occupation & -0.3105 & -0.0947 & -0.2263 & -0.2737 & -0.0316 & -0.0263 & -0.0263 \\
\quad + skin & -0.3105 & -0.1947 & -0.1789 & -0.1895 & -0.1632 & -0.1579 & -0.1579 \\
\quad + skin, hair color, hair type & -0.2368 & -0.2053 & -0.1895 & -0.1947 & -0.2053 & -0.2053 & -0.2053 \\
\midrule
gender + skin (no occ.) & -0.1053 & -0.1474 & -0.1474 & -0.1526 & -0.1579 & -0.1579 & -0.1579 \\
gender + hair color (no occ.) & -0.1211 & -0.0895 & -0.0789 & -0.1053 & -0.0579 & -0.0579 & -0.0579 \\
gender + hair type (no occ.) & -0.2789 & -0.1947 & -0.1947 & -0.2158 & -0.1789 & -0.1789 & -0.1789 \\
skin, hair color, hair type (no occ.) & -0.0421 & -0.2579 & -0.2632 & -0.2579 & -0.2684 & -0.2737 & -0.2737 \\
\midrule
\multicolumn{8}{c}{\textbf{Single-Attribute Differential (T1 $-$ T0)}} \\
gender & & \multicolumn{6}{l}{+0.6211 (AIE = -0.0947)} \\
skin & & \multicolumn{6}{l}{-0.0421} \\
hair color & & \multicolumn{6}{l}{-0.3789} \\
hair type & & \multicolumn{6}{l}{+0.0316} \\
\bottomrule
\end{tabular}
\caption{\textbf{Multi-attribute bias: InstructBLIP-7B.} Per-layer AIE across attribute combinations. Unlike the LLaVA models, InstructBLIP shows no clear language-vision separation: vision layers mediate at comparable or higher levels than language layers, particularly when occupation is absent (see rows without occupation in the table). FBS is consistently high (0.86-0.99), indicating the model almost always responds ``yes'' to factual prompts. 1000 samples, intervention strength 0.4.}
\label{tab:multi_instructblip}
\end{table*}

\section{Uncertainty Quantification and Statistical Significance}
\label{sec:uncertainty}

Tables~\ref{tab:uncertainty_llava}--\ref{tab:uncertainty_minicpm_internvl} present 95\% confidence intervals and Cohen's $d$ effect sizes for AIE and ATE across eight models. Several findings stand out. First, nearly all effect sizes are negative, indicating systematic biases across models and datasets. Second, Cohen's $d$ values are broadly consistent across layers within each model---particularly for vision layers and ATE---suggesting that bias magnitude is substantially a model-level property rather than layer-specific; MiniCPM-V~2.6 shows larger layer-wise variation in language-layer AIE, reflecting its architectural distinctiveness. Third, MS COCO yields substantially larger effects (often $|d| > 1.0$) than FACET, pointing to dataset-dependent bias severity. Notably, several near-zero or small positive AIE values appear on FACET---InstructBLIP-7B visual layer~15 ($d = 0.033$), InstructBLIP-13B visual layer~15 ($d = 0.042$), and LLaVA-NeXT-13B language layer~0 ($d = 0.045$)---where the bias direction effectively reverses; in all three cases the confidence interval spans zero, indicating the reversal is not statistically significant. ATE remains negative for all three models. Across model families, scaling model size does not consistently reduce effect magnitudes, confirming that mechanistic diagnosis reveals persistent fairness-relevant signals.

\begin{table*}[t]
\centering
\setlength{\tabcolsep}{1pt}
\footnotesize
\begin{tabular}{ccc*{8}{c}}
\toprule
\multirow{4}{*}{\textbf{Model}}&\multirow{4}{*}{\textbf{Type}}&\multirow{4}{*}{\textbf{Layer}}&\multicolumn{4}{c}{\textbf{FACET}}&\multicolumn{4}{c}{\textbf{MS COCO}}\\
\cmidrule(lr){4-7} \cmidrule(lr){8-11}
\multicolumn{3}{c}{} &\multicolumn{2}{c}{\textbf{AIE}}&\multicolumn{2}{c}{\textbf{ATE}}&\multicolumn{2}{c}{\textbf{AIE}}&\multicolumn{2}{c}{\textbf{ATE}}\\
\cmidrule(lr){4-5} \cmidrule(lr){6-7} \cmidrule(lr){8-9} \cmidrule(lr){10-11}
&&&\multicolumn{1}{c}{\textbf{95\% CI}}&\multicolumn{1}{c}{\textbf{Cohen\textquotesingle s d}}&\multicolumn{1}{c}{\textbf{95\% CI}}&\multicolumn{1}{c}{\textbf{Cohen\textquotesingle s d}}&\multicolumn{1}{c}{\textbf{95\% CI}}&\multicolumn{1}{c}{\textbf{Cohen\textquotesingle s d}}&\multicolumn{1}{c}{\textbf{95\% CI}}&\multicolumn{1}{c}{\textbf{Cohen\textquotesingle s d}}\\
\midrule
\multirow{6}{*}{L7b} & \multirow{3}{*}{Lang.} & 0 &[-0.047, -0.030]&-0.277&[-0.131, -0.100]&-0.450&[-0.066, -0.047]&-0.353&[-0.249, -0.216]&-0.856 \\
 && 16 &[-0.079, -0.058]&-0.398&[-0.131, -0.100]&-0.450&[-0.072, -0.051]&-0.366&[-0.249, -0.216]&-0.856 \\
 && 31 &[-0.087, -0.064]&-0.421&[-0.131, -0.100]&-0.450&[-0.075, -0.053]&-0.373&[-0.249, -0.216]&-0.856 \\ \cmidrule(lr){2-11}
 &\multirow{3}{*}{Vis.}& 0 &[-0.047, -0.031]&-0.291&[-0.131, -0.100]&-0.450&[-0.064, -0.045]&-0.343&[-0.249, -0.216]&-0.856 \\
 && 8 &[-0.048, -0.031]&-0.293&[-0.131, -0.100]&-0.450&[-0.065, -0.046]&-0.346&[-0.249, -0.216]&-0.856 \\
 && 15 &[-0.049, -0.032]&-0.297&[-0.131, -0.100]&-0.450&[-0.065, -0.046]&-0.346&[-0.249, -0.216]&-0.856 \\ \cmidrule(lr){2-11}
\multirow{6}{*}{L13b} &\multirow{3}{*}{Lang.}& 0 &[-0.134, -0.108]&-0.565&[-0.167, -0.130]&-0.496&[-0.213, -0.183]&-0.809&[-0.309, -0.279]&-1.175 \\
 && 16 &[-0.172, -0.144]&-0.679&[-0.167, -0.130]&-0.496&[-0.262, -0.231]&-0.984&[-0.309, -0.279]&-1.175 \\
 && 31 &[-0.180, -0.150]&-0.702&[-0.167, -0.130]&-0.496&[-0.271, -0.239]&-1.020&[-0.309, -0.279]&-1.175 \\ \cmidrule(lr){2-11}
 &\multirow{3}{*}{Vis.}& 0 &[-0.130, -0.103]&-0.551&[-0.167, -0.130]&-0.496&[-0.202, -0.172]&-0.771&[-0.309, -0.279]&-1.175 \\
 && 8 &[-0.129, -0.102]&-0.548&[-0.167, -0.130]&-0.496&[-0.201, -0.171]&-0.769&[-0.309, -0.279]&-1.175 \\
 && 15 &[-0.128, -0.102]&-0.546&[-0.167, -0.130]&-0.496&[-0.202, -0.172]&-0.773&[-0.309, -0.279]&-1.175 \\
\bottomrule
\end{tabular}
\caption{\textbf{Uncertainty quantification and statistical significance results for LLaVA models on FACET and MS COCO.} Abbreviations: L7b = LLaVA-7B, L13b = LLaVA-13B, Lang. = Language, Vis. = Vision. Confidence intervals (CI) are at 95\% level. Cohen\textquotesingle s d effect sizes are reported for both AIE and ATE.}
\label{tab:uncertainty_llava}
\end{table*}

\begin{table*}[t]
\centering
\setlength{\tabcolsep}{1pt}
\footnotesize
\begin{tabular}{ccc*{8}{c}}
\toprule
\multirow{4}{*}{\textbf{Model}}&\multirow{4}{*}{\textbf{Type}}&\multirow{4}{*}{\textbf{Layer}}&\multicolumn{4}{c}{\textbf{FACET}}&\multicolumn{4}{c}{\textbf{MS COCO}}\\
\cmidrule(lr){4-7} \cmidrule(lr){8-11}
\multicolumn{3}{c}{} &\multicolumn{2}{c}{\textbf{AIE}}&\multicolumn{2}{c}{\textbf{ATE}}&\multicolumn{2}{c}{\textbf{AIE}}&\multicolumn{2}{c}{\textbf{ATE}}\\
\cmidrule(lr){4-5} \cmidrule(lr){6-7} \cmidrule(lr){8-9} \cmidrule(lr){10-11}
&&&\multicolumn{1}{c}{\textbf{95\% CI}}&\multicolumn{1}{c}{\textbf{Cohen\textquotesingle s d}}&\multicolumn{1}{c}{\textbf{95\% CI}}&\multicolumn{1}{c}{\textbf{Cohen\textquotesingle s d}}&\multicolumn{1}{c}{\textbf{95\% CI}}&\multicolumn{1}{c}{\textbf{Cohen\textquotesingle s d}}&\multicolumn{1}{c}{\textbf{95\% CI}}&\multicolumn{1}{c}{\textbf{Cohen\textquotesingle s d}}\\
\midrule
\multirow{6}{*}{NeXT7b} &\multirow{3}{*}{Lang.}& 0 &[-0.147, -0.120]&-0.600&[-0.193, -0.154]&-0.553&[-0.163, -0.134]&-0.640&[-0.321, -0.290]&-1.189 \\
 && 16 &[-0.179, -0.150]&-0.697&[-0.193, -0.154]&-0.553&[-0.195, -0.165]&-0.746&[-0.321, -0.290]&-1.189 \\
 && 31 &[-0.181, -0.151]&-0.702&[-0.193, -0.154]&-0.553&[-0.200, -0.170]&-0.761&[-0.321, -0.290]&-1.189 \\ \cmidrule(lr){2-11}
 &\multirow{3}{*}{Vis.}& 0 &[-0.088, -0.065]&-0.425&[-0.193, -0.154]&-0.553&[-0.161, -0.133]&-0.645&[-0.321, -0.290]&-1.189 \\
 && 8 &[-0.087, -0.065]&-0.423&[-0.193, -0.154]&-0.553&[-0.161, -0.133]&-0.645&[-0.321, -0.290]&-1.189 \\
 && 15 &[-0.087, -0.065]&-0.423&[-0.193, -0.154]&-0.553&[-0.161, -0.132]&-0.643&[-0.321, -0.290]&-1.189 \\ \cmidrule(lr){2-11}
\multirow{6}{*}{NeXT13b} &\multirow{3}{*}{Lang.}& 0 &[-0.003, 0.006]&0.045&[-0.248, -0.071]&-0.359&[-0.071, -0.042]&-0.789&[-0.627, -0.491]&-1.633 \\
 && 16 &[-0.140, -0.080]&-0.731&[-0.248, -0.071]&-0.359&[-0.223, -0.167]&-1.385&[-0.627, -0.491]&-1.633 \\
 && 31 &[-0.106, -0.056]&-0.640&[-0.248, -0.071]&-0.359&[-0.211, -0.166]&-1.635&[-0.627, -0.491]&-1.633 \\ \cmidrule(lr){2-11}
 &\multirow{3}{*}{Vis.}& 0 &[-0.084, -0.062]&-0.413&[-0.148, -0.113]&-0.466&[-0.231, -0.200]&-0.870&[-0.323, -0.291]&-1.230 \\
 && 8 &[-0.082, -0.061]&-0.408&[-0.148, -0.113]&-0.466&[-0.232, -0.201]&-0.873&[-0.323, -0.291]&-1.230 \\
 && 15 &[-0.083, -0.061]&-0.410&[-0.148, -0.113]&-0.466&[-0.232, -0.201]&-0.873&[-0.323, -0.291]&-1.230 \\
\bottomrule
\end{tabular}
\caption{\textbf{Uncertainty quantification and statistical significance results for LLaVA-NeXT models on FACET and MS COCO.} Abbreviations: NeXT-7B = LLaVA-NeXT-7B, NeXT-13B = LLaVA-NeXT-13B, Lang. = Language, Vis. = Vision. Confidence intervals (CI) are at 95\% level. Cohen\textquotesingle s d effect sizes are reported for both AIE and ATE.}
\label{tab:uncertainty_llavanext}
\end{table*}

\begin{table*}[t]
\centering
\setlength{\tabcolsep}{1pt}
\footnotesize
\begin{tabular}{ccc*{8}{c}}
\toprule
\multirow{4}{*}{\textbf{Model}}&\multirow{4}{*}{\textbf{Type}}&\multirow{4}{*}{\textbf{Layer}}&\multicolumn{4}{c}{\textbf{FACET}}&\multicolumn{4}{c}{\textbf{MS COCO}}\\
\cmidrule(lr){4-7} \cmidrule(lr){8-11}
\multicolumn{3}{c}{} &\multicolumn{2}{c}{\textbf{AIE}}&\multicolumn{2}{c}{\textbf{ATE}}&\multicolumn{2}{c}{\textbf{AIE}}&\multicolumn{2}{c}{\textbf{ATE}}\\
\cmidrule(lr){4-5} \cmidrule(lr){6-7} \cmidrule(lr){8-9} \cmidrule(lr){10-11}
&&&\multicolumn{1}{c}{\textbf{95\% CI}}&\multicolumn{1}{c}{\textbf{Cohen\textquotesingle s d}}&\multicolumn{1}{c}{\textbf{95\% CI}}&\multicolumn{1}{c}{\textbf{Cohen\textquotesingle s d}}&\multicolumn{1}{c}{\textbf{95\% CI}}&\multicolumn{1}{c}{\textbf{Cohen\textquotesingle s d}}&\multicolumn{1}{c}{\textbf{95\% CI}}&\multicolumn{1}{c}{\textbf{Cohen\textquotesingle s d}}\\
\midrule
\multirow{6}{*}{IB7b} &\multirow{3}{*}{Lang.}& 0 &[-0.015, -0.013]&-0.801&[-0.134, -0.107]&-0.565&[-0.015, -0.013]&-0.899&[-0.222, -0.204]&-1.454 \\
 && 16 &[-0.051, -0.044]&-0.855&[-0.134, -0.107]&-0.565&[-0.058, -0.053]&-1.351&[-0.222, -0.204]&-1.454 \\
 && 31 &[-0.065, -0.056]&-0.851&[-0.134, -0.107]&-0.565&[-0.060, -0.055]&-1.337&[-0.222, -0.204]&-1.454 \\ \cmidrule(lr){2-11}
 &\multirow{3}{*}{Vis.}& 0 &[-0.000, 0.000]&-0.032&[-0.134, -0.107]&-0.565&[-0.000, 0.000]&-0.011&[-0.222, -0.204]&-1.454 \\
 && 8 &[-0.000, 0.000]&-0.045&[-0.134, -0.107]&-0.565&[-0.000, 0.000]&-0.009&[-0.222, -0.204]&-1.454 \\
 && 15 &[-0.000, 0.000]&0.033&[-0.134, -0.107]&-0.565&[-0.000, 0.000]&-0.038&[-0.222, -0.204]&-1.454 \\ \cmidrule(lr){2-11}
\multirow{6}{*}{IB13b} &\multirow{3}{*}{Lang.}& 0 &[-0.011, -0.010]&-1.000&[-0.218, -0.176]&-0.595&[-0.011, -0.009]&-1.270&[-0.382, -0.355]&-1.685 \\
 && 16 &[-0.053, -0.047]&-0.971&[-0.218, -0.176]&-0.595&[-0.046, -0.043]&-1.750&[-0.382, -0.355]&-1.685 \\
 && 31 &[-0.078, -0.068]&-0.970&[-0.218, -0.176]&-0.595&[-0.069, -0.064]&-1.567&[-0.382, -0.355]&-1.685 \\ \cmidrule(lr){2-11}
 &\multirow{3}{*}{Vis.}& 0 &[-0.000, 0.000]&0.019&[-0.218, -0.176]&-0.595&[-0.000, -0.000]&-0.062&[-0.382, -0.355]&-1.685 \\
 && 8 &[-0.000, 0.000]&0.001&[-0.218, -0.176]&-0.595&[-0.000, 0.000]&-0.002&[-0.382, -0.355]&-1.685 \\
 && 15 &[-0.000, 0.000]&0.042&[-0.218, -0.176]&-0.595&[-0.000, 0.000]&-0.049&[-0.382, -0.355]&-1.685 \\
\bottomrule
\end{tabular}
\caption{\textbf{Uncertainty quantification and statistical significance results for InstructBLIP models on FACET and MS COCO.} Abbreviations: IB7b = InstructBLIP-7B, IB13b = InstructBLIP-13B, Lang. = Language, Vis. = Vision. Confidence intervals (CI) are at 95\% level. Cohen\textquotesingle s d effect sizes are reported for both AIE and ATE.}
\label{tab:uncertainty_instructblip}
\end{table*}

\begin{table*}[t]
\centering
\setlength{\tabcolsep}{1pt}
\footnotesize
\begin{tabular}{ccc*{8}{c}}
\toprule
\multirow{4}{*}{\textbf{Model}}&\multirow{4}{*}{\textbf{Type}}&\multirow{4}{*}{\textbf{Layer}}&\multicolumn{4}{c}{\textbf{FACET}}&\multicolumn{4}{c}{\textbf{MS COCO}}\\
\cmidrule(lr){4-7} \cmidrule(lr){8-11}
\multicolumn{3}{c}{} &\multicolumn{2}{c}{\textbf{AIE}}&\multicolumn{2}{c}{\textbf{ATE}}&\multicolumn{2}{c}{\textbf{AIE}}&\multicolumn{2}{c}{\textbf{ATE}}\\
\cmidrule(lr){4-5} \cmidrule(lr){6-7} \cmidrule(lr){8-9} \cmidrule(lr){10-11}
&&&\multicolumn{1}{c}{\textbf{95\% CI}}&\multicolumn{1}{c}{\textbf{Cohen\textquotesingle s d}}&\multicolumn{1}{c}{\textbf{95\% CI}}&\multicolumn{1}{c}{\textbf{Cohen\textquotesingle s d}}&\multicolumn{1}{c}{\textbf{95\% CI}}&\multicolumn{1}{c}{\textbf{Cohen\textquotesingle s d}}&\multicolumn{1}{c}{\textbf{95\% CI}}&\multicolumn{1}{c}{\textbf{Cohen\textquotesingle s d}}\\
\midrule
\multirow{3}{*}{MCPM} &\multirow{3}{*}{Lang.}& 0 &[-0.062, -0.031]&-0.597&[-3.162, -1.107]&-0.412&[-0.250, -0.158]&-0.886&[-5.537, -4.477]&-1.875 \\
 && 13 &[-1.250, -0.814]&-0.938&[-3.162, -1.107]&-0.412&[-2.576, -1.925]&-1.371&[-5.537, -4.477]&-1.875 \\
 && 27 &[-1.005, -0.387]&-0.447&[-3.162, -1.107]&-0.412&[-1.728, -1.396]&-1.865&[-5.536, -4.476]&-1.875 \\ \cmidrule(lr){2-11}
\multirow{3}{*}{IV3.5} &\multirow{3}{*}{Lang.}& 0 &[-0.080, -0.041]&-0.610&[-6.492, -3.382]&-0.630&[-0.055, -0.023]&-0.491&[-5.403, -3.895]&-1.224 \\
 && 16 &[-0.423, -0.204]&-0.568&[-6.492, -3.382]&-0.630&[-0.768, -0.505]&-0.960&[-5.403, -3.895]&-1.224 \\
 && 31 &[-0.838, -0.399]&-0.559&[-6.492, -3.382]&-0.630&[-0.794, -0.583]&-1.300&[-5.403, -3.895]&-1.224 \\
\bottomrule
\end{tabular}
\caption{\textbf{Uncertainty quantification and statistical significance results for MiniCPM-V~2.6 and \mbox{InternVL3.5} on FACET and MS COCO.} Abbreviations: MCPM = MiniCPM-V~2.6, IV3.5 = \mbox{InternVL3.5}, Lang. = Language, Vis. = Vision. Confidence intervals (CI) are at 95\% level. Cohen\textquotesingle s d effect sizes are reported for both AIE and ATE.}
\label{tab:uncertainty_minicpm_internvl}
\end{table*}

\section{Ethical Considerations}
\label{sec:ethical}

\textbf{Societal context of gender bias research.} This work studies gender bias in vision-language models---a problem with documented real-world consequences, from occupational stereotyping in image search to misgendering in automated captioning. By developing tools that reveal \emph{which internal components} contribute to biased output, we aim to enable more principled fairness interventions. However, we recognize that studying gender as a binary variable (male/female)---an operationalization inherited from the FACET dataset---excludes non-binary and transgender identities. This limitation, discussed in the Limitations section above, carries ethical weight: binary framing can reinforce the very categories that fairness research seeks to problematize. We urge dataset creators to develop more inclusive annotation schemes, and we caution readers against interpreting our findings as characterizing gender expression in its full complexity.

\textbf{Dual-use risk.} The causal mediation framework we develop is a diagnostic instrument: it reveals which layers and modalities show the greatest output sensitivity under controlled intervention, and how strongly activation replacement at each component shifts outputs. This same capability could, in principle, be repurposed to design more efficient biased models---for example, by selectively amplifying components identified through layer-wise contribution signatures. Mechanism-level transparency is value-neutral: it can serve fairness or harm it. We believe the benefits of exposing layer-wise output sensitivity---enabling targeted, evidence-based debiasing---outweigh the risk of misuse, but we flag this tension explicitly. Researchers deploying these methods should consider safeguards such as pairing mediation tools with documented debiasing protocols.

\textbf{Computational and environmental cost.} Causal mediation analysis on large vision-language models carries non-trivial computational costs. Our method requires forward passes under multiple counterfactual conditions for each layer, each prompt, and each model. For the full layer-wise analysis across six core models plus two supplementary architectures reported in this paper, the total GPU-hours are substantial. As mechanism-level evaluation scales to larger models and broader benchmarks, the environmental footprint will grow. We encourage the community to develop more sample-efficient mediation estimators (see our ablation on sampling density, Section~\ref{sec:experiments}) and to report computational budgets alongside mediation results, following best practices for Green NLP.

\textbf{Responsible deployment of diagnostic findings.} Our central finding---that under controlled intervention, language-layer activations exhibit the greatest output sensitivity, with the direct component often carrying the opposite sign under this decomposition---carries implications for fairness engineering. A naive reading might suggest ``removing'' or ``zeroing out'' language layers with large AIE magnitudes. We strongly caution against such interpretations. Layer-wise mediation signatures describe \emph{where} activation replacement produces the largest output shift under a specific experimental design; they do not prescribe which components should be modified. Intervening on components identified through mediation analysis without understanding their broader functional role risks degrading model performance on legitimate tasks, potentially harming downstream users who rely on these models for accessibility, translation, or information retrieval. Mechanism-level evaluation should inform intervention design, not replace careful ablation and task-level validation.

\textbf{Data and model transparency.} All models studied in this work are openly available (LLaVA-1.5, LLaVA-NeXT, InstructBLIP, MiniCPM-V~2.6, \mbox{InternVL3.5}), and both FACET and MS COCO are publicly released benchmarks. We will release our code, mediation estimates, and diagnostic outputs to facilitate reproducibility and independent verification. We note that FACET's annotations were produced by human raters under specific demographic and geographic sampling constraints; the resulting labels may not represent global conceptions of gender or occupation. Researchers applying our framework to new datasets should scrutinize annotation provenance and demographic coverage.

\end{document}